\documentclass[12pt, draftclsnofoot, onecolumn]{IEEEtran}
\ifCLASSINFOpdf

\else

\fi
\usepackage[cmex10]{amsmath}
\usepackage{amssymb}
\usepackage{cite}
\usepackage{graphicx}
\usepackage{array,color}
\usepackage{amsmath}
\usepackage{stfloats}
\usepackage{graphicx}
\usepackage{subfigure}
\usepackage{tabularx}
\usepackage{epsfig,epsf,color,balance,cite}
\usepackage{algorithmic}
\usepackage{algorithm}
\usepackage{bm}
\usepackage{textcomp}

\begin{document}

\title{Joint Precoding and RRH selection for User-centric Green MIMO C-RAN}

\author{Cunhua Pan,  Huiling Zhu, Nathan J. Gomes, and Jiangzhou Wang, \emph{Fellow}, \emph{IEEE}
\thanks{This work has been accepted by IEEE TWC. Part of this work has been presented in IEEE Globecom 2016 \cite{Panconference}.}
\thanks{C. Pan was with the School of Engineering and Digital Arts, University of Kent, Canterbury, Kent, CT2 7NZ, U.K. He is now with the School of Electronic Engineering and Computer Science, Queen Mary University of London, London E1 4NS, U.K. (Email:\{C.Pan\}@qmul.ac.uk)}
\thanks{H. Zhu, N. Gomes and J. Wang are with the School of Engineering and Digital Arts, University of Kent, Canterbury, Kent, CT2 7NZ, U.K. (Email:\{ H.Zhu, N.J.Gomes, J.Z.Wang\}@kent.ac.uk).}
}

\maketitle

\begin{abstract}
This paper jointly optimizes the precoding matrices and the set of active remote radio heads (RRHs) to minimize the network power consumption (NPC) for a user-centric cloud radio access network (C-RAN), where  both the RRHs and users have multiple antennas and each user is served by its nearby RRHs. Both  users' rate requirements and  per-RRH power constraints are considered. Due to these conflicting constraints, this optimization problem may be infeasible. In this paper, we propose to solve this problem in two stages. In Stage I, a low-complexity user selection algorithm is proposed to find the largest subset of feasible users. In Stage II, a low-complexity algorithm is proposed to solve the optimization problem with the users selected from Stage I. Specifically,  the re-weighted $l_1$-norm minimization method is used to transform the original problem with non-smooth objective function into a series of weighted power minimization (WPM) problems, each of which can be solved by the weighted minimum mean square error (WMMSE) method. The solution obtained by the WMMSE method is proved to satisfy the Karush-Kuhn-Tucker (KKT) conditions of the WPM problem. Moreover, a low-complexity algorithm based on Newton's method and the gradient descent method is developed to update the precoder matrices in each iteration of the WMMSE method. Simulation results demonstrate the rapid convergence of the proposed algorithms and the benefits of equipping multiple antennas at the user side. Moreover, the proposed algorithm is shown to achieve near-optimal performance in terms of NPC.
\end{abstract}

\IEEEpeerreviewmaketitle

\section{Introduction}
Mobile communications has been developing very rapidly \cite{huilingtcom09,huilingtcom12,huiling12jsac}.  In recent years, C-RAN has been proposed as a promising solution to support the exponential growth of mobile data traffic  \cite{mobile2011c,alliance2013suggestions}. In C-RAN, all the baseband processing is performed at the baseband unit (BBU) pool with powerful computation capacity, while the remote radio heads (RRHs)  perform the basic functionalities of signal processing \cite{Huiling2011,jiangzhou2012}. The RRHs are geographically distributed away from each other, but connected to the BBU pool through   optical fiber transport links. Under the C-RAN architecture, centralized signal processing technologies can be realized. Hence, significant performance gains can be achieved. In addition,  the RRHs can be densely deployed in the network with low operation cost due to their simple functionalities. This will significantly reduce the average access distance for the users, and thus lowers the transmission power.

On the other hand, it was reported that the total energy consumption of  wireless communications contributes more than 3 percent of the worldwide electrical
energy consumption \cite{fettweis2008ict}, and this portion is expected to grow in the near future due to the explosive growth of high-data-rate applications and mobile devices. Hence, energy efficiency  has attracted extensive interest and becomes one of the main performance metrics in the future fifth generation (5G) systems \cite{Andrews2014,Panjsac,pantwc,panwcl,renhon}. When a large number of RRHs  are deployed in the network, the network power consumption (NPC) of  C-RAN will become considerable due to the increasing circuit power consumption of the RRHs. Fortunately, it was reported in \cite{Correia} that the traffic load varies substantially over both time and space due to  user mobility and varying channel state. Hence, the NPC can be significantly reduced by  putting some  RRHs with light load into sleep mode while maintaining the quality of service (QoS) requirements of the users, which is the focus of this paper.

Recently, the NPC minimization problem for C-RAN has been extensively studied in \cite{Yong2013,Ramamonjison2014,Ng2015,Fuxin2014,Ha2015,tao2015content,Yuanming2014,Shixin2015,Jilei2015,Jianhua2015,dai2016energy}. These papers formulated the joint RRH selection and beamforming vector optimization problem as a mixed-integer non-linear programming (MINLP) problem, which has a nonconvex discontinuous $l_0$-norm in the objective function or constraints. We summarize the existing approaches to solve the  MINLP problem as follows. The first approach was proposed in \cite{Yong2013}, which first reformulated the problem as an extended mixed integer second-order cone programming (SOCP) and then applied the branch-and-cut method  to obtain the optimal solution. In the second approach in \cite{Ramamonjison2014,Ng2015}, the MINLP was first decomposed into a master problem and a beamforming subproblem. Then, an iterative algorithm based on the Benders decomposition was derived to find the optimal solution. Although these two approaches yield the optimal solution, they have an exponential complexity. The third approach is the smooth function method, where the $l_0$-norm was approximated as Gaussian-like function in \cite{Fuxin2014}, the exponential function in \cite{Ha2015}, and arctangent function in \cite{tao2015content}. However, the smooth function cannot produce sparse solutions  in general. The last approach was inspired by the compression sensing, named re-weighted $l_1$-norm minimization method \cite{candes2008enhancing}. This method has been widely adopted in the literature \cite{Jian2013,Yuanming2014,Shixin2015,Jilei2015,Jianhua2015,dai2016energy} due to its low computational complexity and sparsity guarantee, which will also be applied in this paper.

All of the above papers only considered the single-antenna user (SAU) case. With the increasing development in antenna technology \cite{huang2014design,Kun2015}, it is possible to equip the wireless devices with multiple antennas. When both the transmitter  and  the receiver are equipped with multiple antennas, multiple streams can be transmitted simultaneously, rather than only one stream in the SAU case.  Simulation results show that with the increasing number of receive antennas, more users can be admitted. Therefore, in this paper, we consider the multiple-antenna user (MAU) case and jointly optimize the precoding matrices and the set of active RRHs  to minimize the NPC subject to users' rate requirements and per-RRH power constraints.

Unfortunately, the techniques in \cite{Yong2013,Ramamonjison2014,Ng2015,Fuxin2014,Ha2015,tao2015content,Yuanming2014,Shixin2015,Jilei2015,Jianhua2015,dai2016energy} dealing with the SAU case cannot be extended directly to the MAU case. The reasons are as follows. Firstly, since the rate constraints and power constraints are conflicting with each other, this problem may be infeasible. In the SAU networks, the rate requirements can be equivalently represented as signal-to-interference-plus-noise ratio (SINR) constraints, which can be transformed into an SOCP problem. Hence, the feasibility of the original problem can be easily checked by solving the SOCP feasibility problem. However, the rate constraints in the MAU case is non-convex and much more complex due to the complicated rate expression, which cannot be transformed into the SOCP formulation as in the SAU case. Hence, new techniques need to be developed to check the feasibility of the original problem. Secondly, even though the original problem is checked to be feasible, how to solve it is still difficult, since it cannot be transformed into an SOCP problem as in the SAU case.  \cite{Qingjiang2011} proposed the weighted minimum mean square error (WMMSE) method to solve the rate maximization problem for MIMO interfering broadcast channels, where the rate expression is in the objective function. Recently, there have been some work in applying the WMMSE method to solve the energy efficiency (measured in bit/s/Joule) optimization problems under rate constraints \cite{Shiwen2014,Yang2015}. However, these researches have not addressed the feasibility problem due to the incorporated rate constraints. Only in \cite{Yang2015},  a heuristic method was proposed to check the feasibility based on the interference alignment technique, under the assumption that the transmit power is approaching infinity, which in not practical. Since the problem considered in this paper imposes power constraints at each RRH, the heuristic method developed in \cite{Yang2015} is not applicable. More importantly, they have not revealed the hidden property of applying WMMSE method to the optimization problem with rate constraints, such as the convergence property and the optimality of the solutions.

To the best of our knowledge, this paper is the first attempt to solve the joint RRH and precoding optimization problem to minimize the NPC in the MAU based user-centric C-RAN, where each user can be served by an arbitrary subset of RRHs.
 Due to the conflicting constraints, this problem may be infeasible. Some users should be removed or rescheduled for the next transmission to guarantee the  rate requirements of other users. We provide a comprehensive analysis for this problem by considering two stages: user selection in Stage I and algorithm design in Stage II. The main contributions of this paper are summarized as follows:
\begin{enumerate}
  \item In Stage I, a low-complexity user selection approach is proposed to maximize the number of admitted users that can have their QoS requirements satisfied. Specifically, in each step we solve an alternative problem by introducing a series of auxiliary variables. This alternative problem is always feasible. By replacing the rate expression in the constraints with its lower-bound, an iterative algorithm is proposed to solve this problem along with the complexity and convergence analysis of the algorithm. The alternative problem should be solved at most $K$ times, where $K$ is the total number of users. Its complexity is much lower than the optimal exhaustive user selection method that has an exponential complexity. Simulation results show that both algorithms achieve similar performance.

  \item In Stage II, a low-complexity algorithm is proposed to solve the NPC minimization problem with the users selected from Stage I. Specifically, the re-weighted $l_1$-norm minimization method is adopted to convert the non-smooth optimization problem into a series of smooth weighted power minimization (WPM) problems. We again replace the rate expression with its lower-bound and adapt the WMMSE algorithm originally designed for a rate maximization problem to solve the WPM problem. In addition, we strictly prove that when the WMMSE algorithm is initialized with a feasible solution, the sequences of precoder matrices generated in the iterative procedure will finally converge to the Karush-Kuhn-Tucker (KKT) point of the WPM problem.
  \item In each iteration of the WMMSE algorithm, there is a subproblem for the precoder matrices being updated with some other fixed variables. Most existing papers \cite{Shiwen2014,Yang2015,Binbinicc,Jianhua2015,Binbin2014} directly transform it into an SOCP problem  and apply the interior point method \cite{boyd2004convex} to solve it, which may incur high computational complexity. In this paper, we go one step further and develop a low-complexity algorithm to solve this subproblem by exploiting its special structure. Specifically, we equivalently solve its dual problem because the subproblem is a convex problem. Fortunately, the objective function of the dual problem is differentiable, and the block coordinate descent (BCD) method is adopted to solve the dual problem. In each iteration of the BCD method,  Newton's method and the gradient descent method are applied to update the Lagrangian multipliers.  It is strictly proved that  the BCD method can obtain the globally optimal solution of the subproblem. Complexity analysis in conjunction with the simulation results show that the BCD method has a much lower computational complexity than the interior point method.
\end{enumerate}

This paper is organized as follows. In Section \ref{system}, we introduce the system model and formulate the optimization problems. In Section \ref{phase1}, a new approach is introduced to select the maximum number of admitted users. An iterative algorithm with low complexity is provided in Section \ref{phase2}. Simulation results are presented in Section \ref{simulation}. Conclusions are drawn in Section \ref{conclusion}.

\textbf{Notations}: Uppercase and lowercase boldface denote matrices and vectors, respectively. For a matrix ${\bf{A}}$, ${\left\| {\bf{A}} \right\|_F}$ denotes the Frobenius norm of   ${\bf{A}}$ and ${{\bf{A}}^H}$ represents the Hermitian transpose of ${\bf{A}}$.  ${\bf{I}}_{m}$ denotes a $m\times m$ identity matrix.  For a vector ${\bf{a}}$, ${\rm{diag}}(\bf{a})$ denotes the diagonal matrix with diagonal elements given by $\bf{a}$.   ${\rm{blkdiag}}(\bf{\cdot})$  represent the block-diagonal matrices. $\mathbb{E}(\cdot)$, and  ${\rm{Tr}}(\cdot)$ represent expectation, trace operators, respectively. ${\bf{A}}\succeq{\bf{B}}$ means ${\bf{A}} - {\bf{B}}$ is a positive semidefinite matrix. For vector ${\bf{a}} \in {\mathbb{C}^{n \times 1}}$, ${\left\| {\bf{a}} \right\|_2}$ is the Euclidean norm. ${\cal C}{\cal N}\left( {{\bf{0}},\sigma^2 {{\bf{I}}}} \right)$ represents the complex circularly symmetric Gaussian distribution with zero mean vector and covariance matrix $\sigma^2 {{\bf{I}}}$. For a vector ${\bf{x}}$, ${\left\| {\bf{x}} \right\|_0}$ is ${l_0}$-norm, means the number of nonzero entries in a vector.

\section{System Model and Problem Formulation}\label{system}

\subsection{System model}

\begin{figure}
\centering
\includegraphics[width=3.0in]{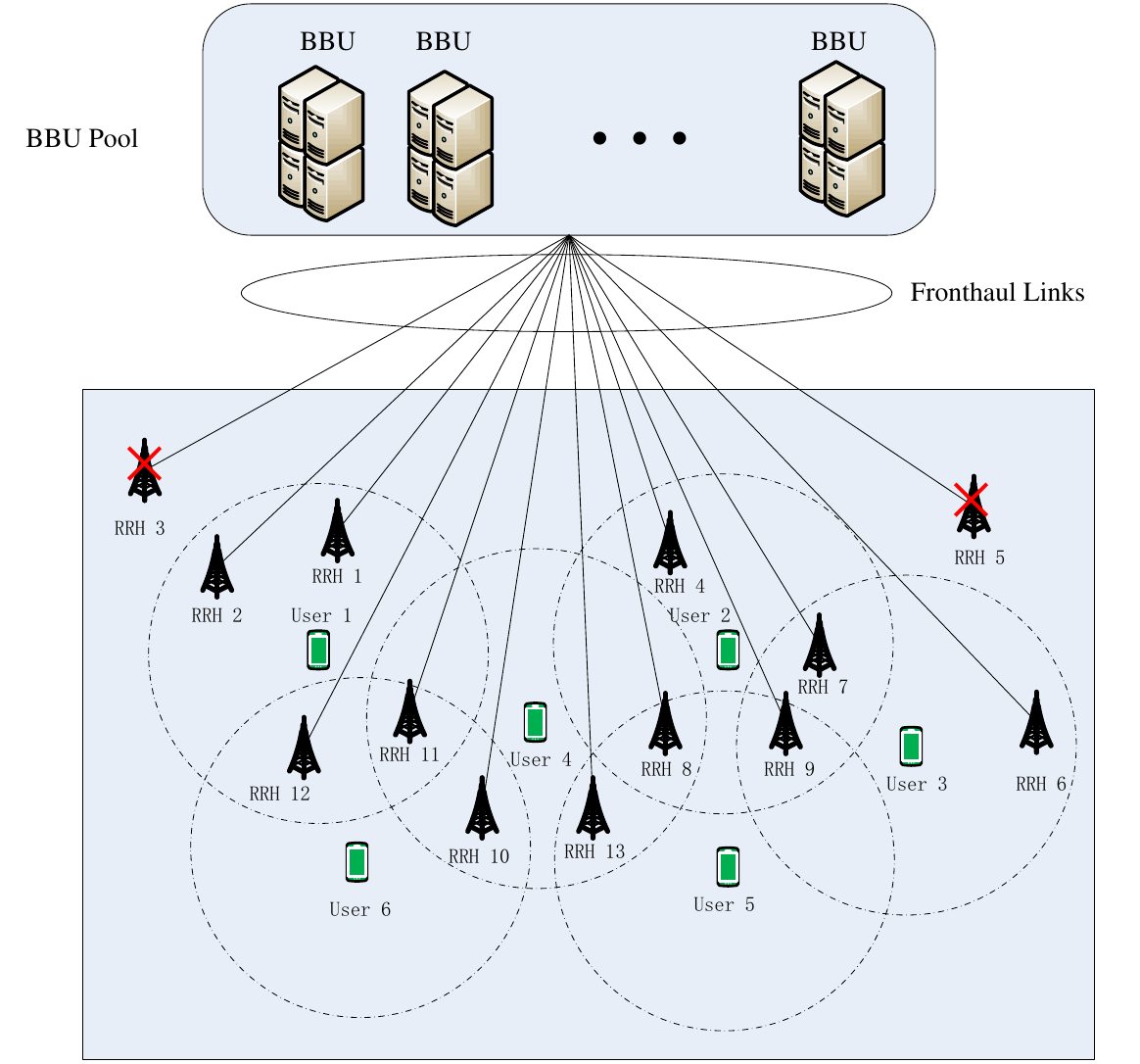}
\caption{Illustration of a C-RAN with thirteen RRHs and six users, where user-centric clustering technique is adopted. In this example, each user is served by its nearby RRHs within the dotted circle centered at itself. The RRHs that are not in any users' candidate set are turned into idle mode, such as RRH3 and RRH 5. }
\label{fig1}
\end{figure}

Consider a downlink C-RAN consisting of $I$ RRHs and $K$ users \footnote{In dense networks, the number of RRHs may be larger than the number of users so that the average distance between serving RRHs and users can be significantly reduced, leading to improved performance. In some extreme cases, each user may be served by its dedicated RRHs as in \cite{Nie2016,Feng2015}, where each RRH serves only one user.}, where each RRH is equipped with $M$ transmit antennas and each user has  $N$ receive antennas, as shown in Fig.~\ref{fig1}. Denote the set of RRHs and users as ${\cal I} = \left\{ {1, \cdots ,I} \right\}$ and $\bar {\cal U} = \left\{ {1, \cdots ,K} \right\}$, respectively.   It is assumed that each RRH is connected to the BBU pool via fronthaul link and the BBU pool has access to all users'  CSI and data information.

Let ${ {\cal U}} \subseteq \bar {\cal U}$  be the set of users that can be admitted to this networks. To reduce the computational complexity of the dense network, the user-centric clustering method is adopted, where each user $k\in {\cal U}$ is assumed to be served by its nearby RRHs since the distant RRHs contribute less to user's signal quality due to the large path loss. The unselected RRHs are turned into idle mode, such as RRH 3 and RRH 5 in Fig.~\ref{fig1}. Let ${{\cal I}_k} \subseteq {\cal I}$ and ${{ {\cal U}}_i}\subseteq { {\cal U}} $  be the candidate set of RRHs for serving user $k$ and candidate set of users served by RRH $i$, respectively. Note that the set of RRHs serving the users may overlap with each other. For example, in Fig.~\ref{fig1}, RRH 12 jointly serves user 1 and user 6.

Denote  ${\bf{V}}_{i,k}\in \mathbb{C}^{M\times d}$ as the precoding matrix used by the $i$th RRH to transmit data vector ${\bf{s}}_k \in \mathbb{C}^{ d \times 1}$  to the $k$th user, where $d$ is the number of data streams for each user, and ${\bf{s}}_k$ satisfies
$\mathbb{E}\left[ {{{\bf{s}}_k}{\bf{s}}_k^{\rm{H}}} \right] = {\bf{I}}_d$ and $\mathbb{E}\left[ {{{\bf{s}}_k}{\bf{s}}_l^{\rm{H}}} \right] = {\bf{0}}, {\rm{for}}\  l \ne k  $. Let ${{\bf{\bar V}}_k} = {\left[ {{\bf{V}}_{i,k}^{\rm{H}},\forall i \in {{\cal I}_k}} \right]^{\rm{H}}} \in {\mathbb{C}^{\left| {{{\cal I}_k}} \right|M \times d}}$ be the big precoding matrix for user $k$ from all RRHs in ${\cal I}_k$. In addition, define a set of new channel matrices ${\bf{\bar H}}_{j,k} = [ {{\bf{H}}_{i,k},\forall i \in {{\cal I}_j}} ] \in {{\mathbb C}^{N \times \left| {{{\cal I}_l}} \right|M}}$, representing the overall CSI from RRHs in ${{\cal I}_j}$ to user $k$, where ${{{\bf{H}}_{i,k}}}\in {\mathbb{C}^{N \times M}}$ denotes the channel matrix from the $i$th RRH to the $k$th user. Then, the received signal vector at the $k$th user, denoted as ${\bf{y}}_k\in \mathbb{C}^{N\times 1} $,  is given by
 \begin{equation}
   \begin{array}{l}
      {\bf{y}}_k={{\bf{\bar H}}_{k,k}}{{\bf{\bar V}}_k}{{\bf{s}}_k}+\sum\nolimits_{j\in {\cal U}, j\neq k}{{\bf{\bar H}}_{j,k}}{{\bf{\bar V}}_j}{{\bf{s}}_j}+{\bf{n}}_k,
   \end{array}
   \end{equation}
where  ${\bf{n}}_k$ is the noise vector at the $k$th user, which satisfies ${\cal C}{\cal N}\left( {{\bf{0}},\sigma  _k^2{{\bf{I}}_N}} \right)$. Then, the achievable rate (nat/s/Hz) of the $k$th user is given by\cite{cover2012elements}
\begin{equation}\label{data}
   \begin{array}{l}
      {R_k}({\bf{V}}) = {\rm{lo}}{{\rm{g}}}\;\left| {{\bf{I}} + {{\bf{\bar H}}_{k,k}}{{\bf{\bar V}}_k}{\bf{\bar V}}_k^H{\bf{\bar H}}_{k,k}^H{\bf{J}}_k^{ - 1}} \right|,
   \end{array}
 \end{equation}
where ${\rm{log}}(\cdot)$ is the base of natural logarithm, ${\bf{J}}_k=\sum\nolimits_{j\in {\cal U}, j\neq k}{{\bf{\bar H}}_{j,k}} {{\bf{\bar V}}_j}{{\bf{\bar V}}_j^H}{{\bf{\bar H}}_{j,k}^H}+{\sigma ^2_k}\bf{I}$ is the interference-plus-noise covariance matrix, and ${\bf{V}}$ is the collection of all precoding matrices. Each user's data rate should be larger than the minimum requirement:
\begin{equation}\label{ratere}
  {\rm{C1}:}\  {R_k}({\bf{V}})\ge R_{k,{\rm{min}}}, \forall k\in {\cal U}.
\end{equation}

With densely deployed RRHs, the power consumption on the RRHs and the corresponding fronthaul links may be significant. Switching off some  RRHs and the corresponding fronthual links  may be a good option to reduce the NPC. To this end, it is critical to model the NPC.

\subsection{NPC model}

The realistic NPC model should consist of three parts: power consumption at the RRHs, that at the fronthaul links and that at the BBU pool.

As in \cite{Yuanming2014}, the power consumption at RRH $i$ can be modeled as follows:
\begin{equation}\label{rrhpower}
  P_i^{{\rm{rrh}}}({\bf{V}})= \left\{ \begin{array}{l}
{{{\eta _i}}}P_i^{{\rm{tr}}}({\bf{V}})+ MP_i^{{\rm{a,rrh}}},\ {\rm{if  }}\ P_i^{{\rm{tr}}}({\bf{V}}) > 0\\
MP_i^{{\rm{s,rrh}}},\qquad \qquad\quad\     {\rm{if}}\ P_i^{{\rm{tr}}}({\bf{V}}) = 0
\end{array} \right.
\end{equation}
where $\eta _i>1$ accounts for the inefficiency of the power amplifier of RRH $i$, $P_i^{{\rm{tr}}}({\bf{V}})$ is the total transmit power of RRH $i$ given by $P_i^{{\rm{tr}}}({\bf{V}}){\rm{ = }}\sum\nolimits_{k \in {{\cal U}_i}} {\left\| {{{\bf{V}}_{i,k}}} \right\|_F^2} $ that satisfies the power constraint:
\begin{equation}\label{powercons}
  {\rm{C2}:}\  P_i^{{\rm{tr}}}({\bf{V}})\le P_{i,\rm{max}}, \forall i\in \cal I,
\end{equation}
$P_i^{{\rm{a,rrh}}}$ and $P_i^{{\rm{s,rrh}}}$ represent the power consumption for each antenna (or each RF chain) when RRH $i$ is in active mode and sleep mode, respectively. In practical systems, $P_i^{{\rm{active}}}$ is much higher than $P_i^{{\rm{sleep}}}$, which motivates us to switch off some RRHs.

In general, more power consumption will be consumed on the fronthaul links when they support high data rates. In \cite{dai2016energy}, this power was modeled to be proportional to the total fronthaul data rate. We modify the model in \cite{dai2016energy} to account for the power when the fronthaul links are in the sleep mode  as follows:
\begin{equation}\label{dieh}
  P_i^{{\rm{fr}}}({\bf{V}}) = \left\{ \begin{array}{l}
{\rho _i}\sum\nolimits_{k \in {{\cal U}_i}} {R_k}({\bf{V}})  + P_i^{a,{\rm{fr}}},\ {\rm{  if  }}\ P_i^{{\rm{tr}}}({\bf{V}}) > 0, \\
P_i^{s,{\rm{fr}}},\qquad \qquad\qquad\quad\quad\    {\rm{ if  }}\ P_i^{{\rm{tr}}}({\bf{V}}) = 0{\rm{.}}
\end{array} \right.
\end{equation}
where $\rho _i$ is the proportional factor for fronthaul link $i$.
The power consumed in the BBU pool mainly depends on the computational complexity for signal processing. However, how to accurately model this kind of power consumption is still not fully understood. As in most papers \cite{Yong2013,Yuanming2014,Shixin2015,dai2016energy}, the BBU power consumption is  modeled as a constant ${P_{{\rm{BBU}}}}$ for simplicity. Let ${\cal A}$ denote the active RRH set. Then, the NPC can be modeled as
\begin{eqnarray}
\hat P({\cal A},{\bf{V}}) &=& \sum\nolimits_{i \in {\cal I}} {\left( {P_i^{{\rm{rrh}}}({\bf{V}}) + P_i^{{\rm{fr}}}({\bf{V}})} \right)}  + {P_{{\rm{BBU}}}}\\
 &=& \sum\nolimits_{i \in {\cal A}} {\left( {{\eta _i}P_i^{{\rm{tr}}}({\bf{V}}) + {\rho _i}\sum\nolimits_{k \in {{\cal U}_i}} {{R_k}({\bf{V}})}  + P_i^c} \right)}  + \sum\nolimits_{i \in {\cal I}} {P_i^s}  + {P_{{\rm{BBU}}}},\label{secondterm}
\label{totalmodel}
\end{eqnarray}
where $P_i^c$ and $P_i^s$ are two constants, given by $P_i^c = M(P_i^{a,{\rm{rrh}}} - P_i^{s,{\rm{rrh}}}) + P_i^{a,{\rm{fr}}} - P_i^{s,{\rm{fr}}}$ and $P_i^s = MP_i^{s,{\rm{rrh}}} + P_i^{s,{\rm{fr}}}$.

%
%

\subsection{Problem Formulation}

Due to the power constraints C2, the rate requirements C1 may not be satisfied for all users. Some users should be removed to make the optimization problem feasible. Hence, we formulate a two-stage optimization problem. In Stage I, one should maximize the number of admitted users that can be supported by the system; in Stage II, one should jointly select some RRHs and optimize the precoding matrices to minimize the NPC with the selected users from Stage I.

Specifically, the optimization problem in Stage I can be formulated as
\begin{equation}\label{staone}
\begin{array}{l}
\mathop {\max }\limits_{{\bf{V}}, {\cal U} \subseteq {\overline {\cal U}} } \quad \left| \cal U \right|\\
\quad {\rm{s}}.{\rm{t}}.\quad {\kern 1pt} {\rm{C1}},{\rm{C2}}.
\end{array}
\end{equation}
Then in Stage II, we aim to jointly select the RRHs and optimize the precoding matrices to minimize the NPC with the users selected from Stage I, which can be formulated as\footnote{In general, the number of transmit antennas should be optimized to additionally reduce the NPC as seen in the RRH power consumption model in (\ref{rrhpower}). However, the resulting problem will be much more difficult to solve, and will be left for future work.}
\begin{subequations}\label{mainpro}
\begin{align}
\mathop {\min }\limits_{{\cal A},{\bf{V}}} \quad
& \sum\nolimits_{i \in {\cal A}} {\left( {{\eta _i}P_i^{{\rm{tr}}}({\bf{V}}) + {\rho _i}\sum\nolimits_{k \in {{\cal U}_i^{\star}}} {{R_k}({\bf{V}})}  + P_i^c} \right)}   \label{objfunct}
\\
\qquad\ \textrm{s.t.}\qquad\!\!\!\!
&\  {\rm{C1}}, \sum\nolimits_{k \in {{\cal U}_i^{\star}}} {\left\| {{{\bf{V}}_{i,k}}} \right\|_F^2}  \le {P_{i,\max }}, i \in {\cal A},\\
&\   \sum\nolimits_{k \in {{\cal U}_i^{\star}}} {\left\| {{{\bf{V}}_{i,k}}} \right\|_F^2}=0,  i\in  {\cal I}\backslash {\cal A},
\end{align}
\end{subequations}
where ${\cal U}_i^{\star}$ is the solution from Stage I. Note that when the system parameters are given, the last two terms in (\ref{secondterm}) are constants, and are omitted in the objective function.

Both the optimization problems in the two stages are MINLP problems and are difficult to solve.  The intuitive approach to solve this kind of problems is through the exhaustive search. For example, to solve the NPC minimization problem in Stage II, one must solve the precoding matrices that minimizes the NPC with each given  $\cal A$ and obtain the corresponding objective value. Finally, the  $\cal A$ that achieves the minimum NPC together with the corresponding precoding matrices is the optimal solution of Problem (\ref{mainpro}). However,  the exhaustive search has exponentially prohibitive complexity with respect to the number of RRHs, which is hard to be implemented in practice in dense C-RANs. The same issue holds for the user selection problem in Stage I, where the exhaustive search method has an exponential complexity of the number of users. Hence, this motivates us to develop low-complexity algorithms to solve these two Problems.

\section{Stage I: Low-complexity User Selection Algorithm}\label{phase1}

In this section, we provide a low-complexity user selection algorithm to guarantee the rate requirements of other users. Specifically,  for an arbitrary given subset of users $\cal U$, we construct an alternative problem by introducing a series of  auxiliary variables $\{\alpha_k\}_{k\in \cal U}$:
\begin{equation}\label{alternativepro}
 \begin{array}{*{20}{l}}
{\mathop {{\rm{min}}}\limits_{{{\left\{ {{\alpha _k}} \right\}}_{k \in {\cal U}}},{\bf{V}}} \;\sum\nolimits_{k \in {\cal U}} {{{\left( {{\alpha _k} - 1} \right)}^2}} \;\;}\\
{\;{\rm{s}}.{\rm{t}}.\;\;\;{\rm{C2}},{R_k}({\bf{V}}) \ge \alpha _k^2{R_{k,{\rm{min}}}},\forall k \in {\cal U},}\\
\end{array}
\end{equation}
Obviously, Problem (\ref{alternativepro}) is always feasible and the optimal $\alpha _k$ for each user $k$ should be no larger than one. This can be easily proved by contradiction. Moreover, user $k$ can be admitted if and only if the optimal $\alpha_k$ is equal to one. Hence, maximizing the number of admitted users is equal to finding the largest subset of users $\cal U$, in which all $\{\alpha_k\}_{k\in \cal U}$ are equal to one.

Based on the above analysis, we provide a low-complexity user selection (USC) algorithm to solve Problem (\ref{staone}) in Stage I. The main idea  is to remove each user with the least $\alpha_k<1$ in each iteration. It is intuitive since the user with the least $\alpha_k$ has the largest gap to its rate target.

\begin{algorithm}
\caption{USC Algorithm}\label{selctalg}
\begin{algorithmic}[1]
\STATE  Initialize the set of users ${\cal U}=\{1,\cdots,K\}$;
\STATE Given $\cal U$,  solve Problem (\ref{alternativepro}) by  Algorithm  \ref{iterdada} in Subsection \ref{algoeqd1} to obtain $\left\{ {\alpha _k} \right\}_{k \in \cal U}$ and ${{\bf{V}}}$;
 \STATE If $\alpha_k=1, \forall k\in \cal U$,  output ${\bf{V}}$ and ${\cal U}^*\!\!=\!\!\cal U$ for the initialization of Stage II and terminate; Otherwise, find  ${k^*} \!\!=\!\! \arg \mathop {\min }\nolimits_{k\in \cal U} \alpha _k$, remove user $k^*$ and update ${\cal U} = {\cal U}\backslash {k^*}$, go to step 2.
\end{algorithmic}
\end{algorithm}

\vspace{-0.2cm}\subsection{Algorithm to solve Problem (\ref{alternativepro})}\label{algoeqd1}

In step 2 of Algorithm \ref{selctalg}, Problem (\ref{alternativepro})  needs to be solved.  Due to constraints C3 in (\ref{alternativepro}), Problem (\ref{alternativepro}) is a non-convex problem, which is difficult to solve. To handle this difficulty, we  apply the relationships between WMMSE and the rate expression.

We consider the linear receiver filter so that the estimated signal vector is given by
\begin{equation}\label{decoded}
{{\bf{\hat s}}_k} = {\bf{U}}_k^H{{\bf{y}}_k},\forall k\in \cal U.
\end{equation}
where ${\bf{U}}_k\in \mathbb{C}^{N \times d}$ is the receiver filter of the $k$th user. Since the signal vectors ${\bf{s}}_k$'s and noise ${\bf{n}}_k$'s are mutually independent, the mean square error (MSE) matrix at the $k$th user is given by
 \begin{eqnarray}
{{\bf{E}}_k} &=& {\mathbb{E}_{{\bf{s,n}}}}\left[ {\left( {{{{\bf{\hat s}}}_k} - {{\bf{s}}_k}} \right){{\left( {{{{\bf{\hat s}}}_k} - {{\bf{s}}_k}} \right)}^H}} \right]\nonumber\\
 &=&\!\!\!\!\! \left( {{\bf{ U}}_k^H{{\bf{\bar H}}_{k,k}}{{\bf{\bar V}}_k} \!\!-\!\! {{\bf{I}}_d}} \right){\left( {{\bf{U}}_k^H{{\bf{\bar H}}_{k,k}}{{\bf{\bar V}}_k}\!\! -\!\! {{\bf{I}}_d}} \right)^H}
 \!\!+\!\! \sum\nolimits_{j\in {\cal U}, j\ne k} \!\!{{\bf{U}}_k^H{{\bf{\bar H}}_{j,k}}{{\bf{\bar V}}_j}} {\bf{\bar V}}_j^H{\bf{\bar H}}_{j,k}^H{{\bf{U}}_k} \!\!+ \!\! \sigma_k^2{\bf{U}}_k^H{{\bf{U}}_k}\label{mse}.
\end{eqnarray}
By introducing a set of auxiliary matrices  $\{{{\bf{W}}_k}\succeq {\bf{0}}\}$, we define the following functions
\begin{equation}\label{hfunction}
{h_k}\left( {\bf{V}},{{\bf{U}}_k,{\bf{W}}_k} \right) =  {\log \left| {{{\bf{W}}_k}} \right| - {\rm{Tr}}\left( {{{\bf{W}}_k}{{\bf{E}}_k}} \right) + d},\forall k.
\end{equation}
where ${\bf{E}}_k$ is the MSE matrix of user $k$ given in (\ref{mse}). The following lemma establishes the relationships between the rate expression and function ${h_k}\left( {\bf{V}},{{\bf{U}}_k,{\bf{W}}_k} \right)$.

\itshape \textbf{Lemma 1  \cite{Qingjiang2011} :}  \upshape ${h_k}\left( {\bf{V}},{{\bf{U}}_k,{\bf{W}}_k} \right)$ is a concave function for each set of the matrices ${\bf{V}}$, ${\bf{U}}_k$ and ${\bf{W}}_k$ when the other two are given. Given ${\bf{V}}$,  ${h_k}\left( {\bf{V}},{{\bf{U}}_k,{\bf{W}}_k} \right)$ is the lower-bound of the data rate ${R_k}({\bf{V}})$ in (\ref{data}). The optimal  ${{\bf{U}}_k, {\bf{W}}_k}$ for ${h_k}\left( {\bf{V}},{{\bf{U}}_k,{\bf{W}}_k} \right)$ to achieve the data rate is given by
\begin{equation}\label{receanu}
  {\bf{U}}_k^{\star} = {\left( {\sum\nolimits_{j\in {\cal U}} {{{\bf{\bar H}}_{j,k}}{\bf{\bar V}}_j{\bf{\bar V}}_j^{H}{\bf{\bar H}}_{j,k}^H}  + \sigma  _k^2{\bf{I}}} \right)^{ - 1}}{{\bf{\bar H}}_{k,k}}{\bf{\bar V}}_k, {\bf{W}}_k^{\star} = {\bf{E}}{_k^{\star -1}},\forall k, \forall k
\end{equation}
where ${\bf{E}}_k^{\star}$ is obtained by plugging the expression of ${\bf{U}}_k^{\star}$ into the $k$th user's MSE in (\ref{mse})
\begin{equation}\label{MSESTAR}
{\bf{E}}_k^{\star} = {\bf{I}}_d - {\bf{\bar V}}_k^{H}{\bf{\bar H}}_{k,k}^H{\left( {\sum\nolimits_{j \in \cal U} {{{\bf{\bar H}}_{j,k}}{\bf{\bar V}}_j{\bf{\bar V}}_j^{H}{\bf{\bar H}}_{j,k}^H}  + \sigma  _k^2{\bf{I}}} \right)^{ - 1}}{{\bf{\bar H}}_{k,k}}{\bf{\bar V}}_k.
\end{equation}
\hfill $\Box$

By replacing the first set of constraints in (\ref{alternativepro})  with its lower-bound ${h_k}\left( {\bf{V}},{{\bf{U}}_k,{\bf{W}}_k} \right)$, we have  the following optimization problem
\vspace{-0.25cm}
\begin{equation}\label{feasiEquivalentpro}
   \begin{array}{*{20}{l}}
{\mathop {{\rm{min}}}\limits_{{{\left\{ {{\alpha _k}} \right\}}_{k \in {\cal U}}},{\bf{V}},{\bf{W}},{\bf{U}}} \;\sum\nolimits_{k \in {\cal U}} {{{\left( {{\alpha _k} - 1} \right)}^2}} \;\;}\\
{\;{\rm{s}}{\rm{.t}}{\rm{.}}\;\;\;{\rm{C2}}, {h_k}\left( {{\bf{V}},{{\bf{U}}_k},{{\bf{W}}_k}} \right) \ge \alpha_k^2 {R_{k,{\rm{min}}}},\forall k \in {\cal U},}
\end{array}
   \end{equation}
where ${\bf{U}}$ and ${\bf{W}}$ are the collection of matrices ${{\bf{U}}_k, \forall k}$ and ${{\bf{W}}_k,\forall k}$, respectively.

To solve Problem (\ref{feasiEquivalentpro}), we apply the block coordinate descent method: given ${\bf{V}}$, update ${\bf{U}}$ and ${\bf{W}}$ by using (\ref{receanu}); update $\{\alpha_k\}_{k\in \cal U}$ and ${\bf{V}}$ with given ${\bf{U}}$ and ${\bf{W}}$. We only need to solve the latter one. Putting the MSE expression in (\ref{mse}) into  constraints C4 in Problem  (\ref{feasiEquivalentpro}) yields
\begin{equation}\label{feasequalprosub}
\begin{array}{*{20}{l}}
{\mathop {\min }\limits_{\{\alpha_k\}_{k\in \cal U},{\bf{V}}} \;\sum\nolimits_{k \in {\cal U}} {{{\left( {{\alpha _k} - 1} \right)}^2}} }\\
\begin{array}{l}
{\rm{s}}.{\rm{t}}.\;\ {\rm{C2}}, {\rm{C5}}: {\rm{Tr}}\left( {{{\left( {{\bf{U}}_k^H{{\bf{\bar H}}_{k,k}}{{\bf{\bar V}}_k} - {{\bf{I}}_k}} \right)}^H}{{\bf{W}}_k}\left( {{\bf{U}}_k^H{{\bf{\bar H}}_{k,k}}{{\bf{\bar V}}_k} - {{\bf{I}}_k}} \right)} \right)\\
\qquad\quad\quad + \sum\nolimits_{j\in {\cal U}, j \ne k} {{\rm{Tr}}} \left( {{\bf{\bar V}}_j^H{\bf{\bar H}}_{j,k}^H{{\bf{U}}_k}{{\bf{W}}_k}{\bf{U}}_k^H{{\bf{\bar H}}_{j,k}}{{\bf{\bar V}}_j}} \right) + {\alpha_k ^2}{R_{k,\min }} \le {t_k},\forall k\in \cal U,
\end{array}
\end{array}
\end{equation}
where  ${t_k} = \log \left| {{{\bf{W}}_k}} \right| + d - \sigma _k^2{\rm{Tr}}\left( {{\bf{U}}_k^H{{\bf{U}}_k}{{\bf{W}}_k}} \right)$.

Without loss of generality, we assume ${\cal U}={\bar {\cal U}}=\{1,\cdots,K\}$ and define the indices of ${{{\cal U}_i}}$ as ${{\cal U}_i} = \{ q_1^i, \cdots ,q_{|{{\cal U}_i}|}^i\} $. Problem (\ref{feasequalprosub}) can be equivalently transformed into the following problem
\begin{equation}\label{socpfea}
  \begin{array}{l}
\mathop {\min }\limits_{\{\alpha_k\}_{k\in \cal U},{\bf{V}}} \;\sum\nolimits_{k \in {\cal U}} {{{\left( {{\alpha _k} - 1} \right)}^2}} \\
{\rm{s}}{\rm{.t}}{\rm{. }}\ {\left\| {{{\bf{x}}_k}} \right\|_2} \le \sqrt {{t_k}} ,\forall k\in \cal U,\\
\quad\ \  {\left\| {{{\bf{y}}_i}} \right\|_2} \le \sqrt {{P_{i.\max }}} ,\forall i\in \cal I,
\end{array}
\end{equation}
where ${{\bf{x}}_k}$ is given by
\[\begin{array}{l}
{{\bf{x}}_k} = \left[ {{\rm{vec}}{{\left( {{\bf{\bar V}}_1^H{\bf{\bar H}}_{1,k}^H{{\bf{U}}_k}{\bf{W}}_k^{1/2}} \right)}^H}, \cdots ,} \right.{\rm{vec}}{\left( {\left( {{{\bf{\bar V}}_{k}^H}{{\bf{\bar H}}_{k,k}^H} {\bf{U}}_k - {{\bf{I}}_k}} \right){\bf{W}}_k^{1/2}} \right)^H},\\
\qquad\qquad\qquad\qquad\qquad \qquad   {\left. {{\rm{ }} \cdots ,{\rm{vec}}{{\left( {{\bf{\bar V}}_K^H{\bf{\bar H}}_{K,k}^H{{\bf{U}}_k}{\bf{W}}_k^{1/2}} \right)}^H},\alpha_k \sqrt {{R_{k,\min }}} } \right]^H}
\end{array}\]
and ${\bf{y}}_i$ is given by
\begin{equation}\label{yi}
  {{\bf{y}}_i} = {\left[ {{\rm{vec}}{{\left( {{{\bf{V}}_{i,q_1^i}}} \right)}^H}, \cdots ,{\rm{vec}}{{\left( {{{\bf{V}}_{i,q_{|{{\cal U}_i}|}^i}}} \right)}^H}} \right]^H}.
\end{equation}
Problem (\ref{socpfea}) is an SOCP problem for which a globally optimal solution can be obtained by existing techniques such as  interior point method \cite{boyd2004convex}.

Based on the above analysis, the iterative algorithm for solving Problem (\ref{alternativepro}) is formally described in Algorithm \ref{iterdada}.

\itshape \textbf{Theorem 1:}  \upshape Algorithm \ref{iterdada} will converge during the iterative procedure.

\itshape \textbf{Proof:}  \upshape Please see Appendix \ref{prooftheorem1}. \hfill $\Box$

\begin{algorithm}
\caption{Iterative Algorithm}\label{iterdada}
\begin{algorithmic}[1]
\STATE Initialize iterative number $n=1$, the maximum number of iterations $n_{\rm{max}}$.  Initial precoding matrices ${\bf{V}}^{(0)}$  such that the per-RRH power constraints are satisfied. Calculate ${\bf{U}}^{(0)}$ and ${\bf{W}}^{(0)}$ by using (\ref{receanu}) with ${\bf{V}}^{(0)}$;
\STATE With ${\bf{U}}^{(n-1)}$ and ${\bf{W}}^{(n-1)}$, update $\{\alpha_k^{(n)}\}_{k\in \cal U}$ and ${\bf{V}}^{(n)}$ by solving the SOCP problem (\ref{socpfea});
 \STATE Update ${\bf{U}}^{(n)}$ and ${\bf{W}}^{(n)}$ as in (\ref{receanu})  with ${\bf{V}}^{(n)}$;
 \STATE If  $n< n_{\rm{max}}$, set $n \leftarrow n + 1$  and go to step 2. Otherwise, terminate.
\end{algorithmic}
\end{algorithm}

\subsection{Overall complexity to solve Problem (\ref{staone}) in Stage I}\label{comana}
We first analyze the complexity of Algorithm \ref{iterdada} to solve Problem (\ref{alternativepro}). For simplicity, we assume that candidate size for each user is equal, i.e., $\left| {{{\cal I}_k}} \right| = l$, and ${\cal U}=\bar {\cal U}$. In each iteration of Algorithm \ref{iterdada}, the main  complexity lies in step 2, where the SOCP Problem (\ref{socpfea}) is solved. This problem has $2MKld+K$ real  variables, $K$ SOC constraints where each has $2Kd^2+1$ real dimensions, and $I$ SOC constraints where each has $2Md\left| {{{\cal U}_i}} \right|$ real dimensions. From  [page 196, \cite{lobo1998applications}],
the  complexity  is   $O\left( {{{\left( {2MKld + K} \right)}^2}\left( {2{K^2}{d^2} + K + 2Md\sum\nolimits_{i \in {\cal I}} {\left| {{{\cal U}_i}} \right|} } \right)} \right)$, and the total number of iterations required is  $O\left( {\sqrt {I+K} } \right)$. Note that $\sum\nolimits_{i \in {\cal I}} {\left| {{{\cal U}_i}} \right|} {\rm{ = }}\sum\nolimits_{k \in {\cal U}} {\left| {{{\cal I}_k}} \right|}  = Kl$, the total complexity to solve the SOCP Problem (\ref{socpfea}) is given by $O\left( {\sqrt {I + K} {{\left( {2MKld + K} \right)}^2}\left( {2{K^2}{d^2} + K + 2MdKl } \right)} \right).$ Finally,  Algorithm \ref{iterdada} should be run at most $K$ times, then the overall complexity to solve Problem (\ref{staone}) in Stage I is at most ${T_{{\rm{StageI}}}} = O\left( {K\sqrt {I + K} {{\left( {2MKld + K} \right)}^2}\left( {2{K^2}{d^2} + K + 2MdKl} \right)} \right).$

\section{Stage II: A Low-complexity Algorithm to Solve Problem (\ref{mainpro})}\label{phase2}

In this section, we provide a low-complexity algorithm to solve Problem (\ref{mainpro}) with the selected users from Stage I. First, we adopt the  re-weighted $l_1$-norm  method \cite{candes2008enhancing} to transform the original non-smooth optimization problem into a series of WPM problems.  Then, the WPM problem is solved by the WMMSE algorithm. In each iteration of the WMMSE algorithm, there is a subproblem that the precoder matrices should be optimized. We exploit the special structure of the subproblem and develop a low-complexity algorithm to solve it.
\vspace{-0.2cm}
\subsection{Reweighted $l_1$-norm minimization}

For simplicity, the subscript in ${\cal U^{\star}}$ is omitted. It is easy to see that the minimum rate constraints in C1 of Problem (\ref{mainpro}) hold with equality at the optimal point, i.e., ${R_k}({\bf{V}}) = {R_{k,\min }},\forall k$. Then, defining $\tilde P_i^c \buildrel \Delta \over = {\rho _i}\sum\nolimits_{k \in {\cal U}_i^ \star } {{R_{k,\min }}}  + P_i^c$ and using the ${l_0}$-norm, the objective function of Problem (\ref{mainpro}) is equivalent to
$
\sum\nolimits_{i \in {\cal I}} {\left( {{\eta _i}\sum\nolimits_{k \in {{\cal U}_i}} {\left\| {{{\bf{V}}_{i,k}}} \right\|_F^2 + {{\left\| {\sum\nolimits_{k \in {{\cal U}_i}} {\left\| {{{\bf{V}}_{i,k}}} \right\|_F^2} } \right\|}_0}\tilde P_i^c} } \right)} .$
This rewritten expression enables us to apply the  compressive sensing techniques \cite{sriperumbudur2011majorization}, where the non-smooth ${l_0}$-norm objective can often be approximated by a re-weighted ${l_1}$-norm, i.e.,
\begin{equation}\label{reweighted}
 {\left\| {\sum\nolimits_{k \in {{\cal U}_i}} {\left\| {{{\bf{V}}_{i,k}}} \right\|_F^2} } \right\|_0} \approx a _i^{(n)}\sum\nolimits_{k \in {{\cal U}_i}} {\left\| {{{\bf{V}}_{i,k}}} \right\|_F^2},\vspace{-0.2cm}
\end{equation}
where $a _i^{(n)}$ is a weight factor of the $i$th RRH at the $n$th iteration that is iteratively updated as
\begin{equation}\label{weight}
 a_i^{(n)} = \frac{1}{{\sum\nolimits_{k \in {{\cal U}_i}} {\left\| {{\bf{V}}_{i,k}^{(n)}} \right\|_F^2}  + \delta }},\forall i,
\end{equation}
where $\delta $ is a small constant regularization parameter and ${{\bf{V}}_{i,k}^{(n)}}$ is the solution in the $n$th iteration. The above updating rule  shows that those RRHs with lower transmit power in the previous iteration will have larger weights, which force them to be shut off in the future iterative procedure.

By using the approximation in (\ref{reweighted}), we have the following optimization problem that should be solved
in the $n$-th iteration
\begin{equation}\label{subproblem}
\begin{array}{l}
\mathop {\min }\limits_{\bf{V}}\  \sum\nolimits_{i\in \cal I} { \omega_i^{(n-1)} \sum\nolimits_{k \in {{\cal U}_i}} {\left\| {{{\bf{V}}_{i,k}}} \right\|_F^2} }   \\
{\rm{s}}{\rm{.t.}}\quad
 {\rm{C1,C2}},
\end{array}
\end{equation}
where $\omega _i^{(n-1)} = {\eta _i} +{a_i^{(n-1)}}{\tilde P_i^c}.$

Based on the above analysis, the re-weighted $l_1$-norm based (RLN) algorithm to solve Problem (\ref{mainpro}) is given in Algorithm \ref{rlw}. The convergence of the RLN algorithm is proved  in \cite{Jian2013}. In addition, \cite{Jian2013} showed that the RLN algorithm is guaranteed to achieve sparse solutions, while the other smooth approximations cannot produce sparse solutions in general.
\begin{algorithm}
\caption{RLN algorithm}\label{rlw}
\begin{algorithmic}[1]
\STATE Initialize a small enough $\delta $, the iterative number $n=1$, the maximum number of iterations $N_{\rm{max}}$. Initialize ${\bf{V}}^{(0)}$ with the outputs given by Stage I, calculate $\{\omega _{i}^{(0)},\forall i\}$;
 \STATE Given $\{\omega _{i}^{(n-1)},\forall i\}$, solve Problem (\ref{subproblem}) to get ${\bf{V}}^{(n)}$ by using the WMMSE algorithm in Subsection \ref{subproblemsolve};
 \STATE Update $\{\omega _{i}^{(n)},\forall i\}$ with ${\bf{V}}^{(n)}$;
 \STATE If $n\geq N_{\rm{max}}$, terminate.  Otherwise, set $n \leftarrow n + 1$  and go to step 2.
\end{algorithmic}
\end{algorithm}

\subsection{Algorithm to Solve Problem (\ref{subproblem})}\label{subproblemsolve}

For simplicity, the subscript of $\omega_i^{(n-1)}$ in Problem (\ref{subproblem}) is omitted.  The main difficulty in solving Problem (\ref{subproblem}) lies in the rate requirement, which is non-convex.  To handle this difficulty, we again apply the relationship between WMMSE and the rate expression. Based on Lemma 1, we replace the rate constraints in (\ref{subproblem})  with its lower bound ${h_k}\left( {\bf{V}},{{\bf{U}}_k,{\bf{W}}_k} \right)$. Define the indices of ${{\cal I}_k} $ as ${{\cal I}_k} = \left\{ {s_1^k, \cdots ,s_{\left| {\cal I}_k \right|}^k} \right\}$, we have  the following optimization problem
\begin{equation}\label{Equivalentpro}
   \begin{array}{l}
           \mathop {{\rm{min}}}\limits_{{\bf{V,W,U}}}  \sum\nolimits_{k \in {\cal U}} {{\rm{Tr}}\left( {{\bf{\bar V}}_k^H} {{\bf{G}}_k}{{{\bf{\bar V}}}_k} \right)} \\
           \;\text{s.t.}  \;\;\;{h_k}\left( {{\bf{V}},{\bf{U}}_k,{\bf{W}}_k} \right) \ge {R_{k,{\text{min}}}},\forall k\in \cal U,\\
           \quad\quad  \;\sum\nolimits_{k \in {{\cal U}_i}} {\left\| {{{\bf{B}}_{i,k}}{{{\bf{\bar V}}}_k}} \right\|} _F^2 \le {P_{i,\max }},\forall i \in {\cal I},
   \end{array}\vspace{-0.2cm}
   \end{equation}
where ${{\bf{G}}_k}$ and ${{\bf{B}}_{i,k}}$ are both diagonal matrices, given by
\begin{equation}\label{bi}
  {{\bf{G}}_k} = {\rm{blkdiag}}\left\{ {{\omega _{s_1^k}}{{\bf{I}}_{M \times M}}, \cdots ,{\omega _{s_{\left| {{{\cal I}_k}} \right|}^k}}{{\bf{I}}_{M \times M}}} \right\}\vspace{-0.2cm}
\end{equation}
and
\begin{equation}\label{BIJ}
  {{\bf{B}}_{i,k}}{\rm{ = diag}}\left\{ {\overbrace {{{\bf{0}}_{1 \times M}}}^{s_1^k}, \cdots ,\overbrace {{{\bf{1}}_{1 \times M}}}^{s_j^k},\overbrace {{{\bf{0}}_{1 \times M}}}^{s_{j + 1}^k}, \cdots ,\overbrace {{{\bf{0}}_{1 \times M}}}^{s_{\left| {{{\cal I}_k}} \right|}^k}} \right\},\;{\rm{if}}\;s_j^k = i,\forall i \in {\cal I},k \in {\cal U}.\vspace{-0.2cm}
\end{equation}

By solving  Problem (\ref{Equivalentpro}), we can find a solution that satisfies the KKT conditions of Problem  (\ref{subproblem}). To solve it, we again apply the block coordinate descent method.  Matrices ${\bf{U}}$ and ${\bf{W}}$ can be updated with (\ref{receanu}). The remaining task is to update ${\bf{V}}$ with given ${\bf{U}}$ and ${\bf{W}}$. Plugging the MSE expression in (\ref{mse}) into the first set of Problem  (\ref{Equivalentpro}) yields
\begin{equation}\label{equalprosub}
\begin{array}{l}
 \mathop {{\rm{min}}}\limits_{{\bf{V}}} \  \sum\nolimits_{k \in {\cal U}} {{\rm{Tr}}\left( {{\bf{\bar V}}_k^H} {{\bf{G}}_k}{{{\bf{\bar V}}}_k} \right)} \\
{\rm{s}}{\rm{.t}}{\rm{.  }}\ \sum\nolimits_{j \in {\cal U}} {{\rm{Tr}}\left( {{\bf{\bar V}}_j^H{\bf{\bar H}}_{j,k}^H{{\bf{U}}_k}{{\bf{W}}_k}{\bf{U}}_k^H{{{\bf{\bar H}}}_{j,k}}{{{\bf{\bar V}}}_j}} \right)}  \!\!-\!\! {\rm{Tr}}\left( {{{\bf{W}}_k}{\bf{U}}_k^H{{{\bf{\bar H}}}_{k,k}}{{{\bf{\bar V}}}_k}} \right)\!\! -\!\! {\rm{Tr}}\left( {{\bf{\bar V}}_k^H{\bf{\bar H}}_{k,k}^H{{\bf{U}}_k}{{\bf{W}}_k}} \right)\! \le \! {c_k},\forall k\\
\quad\  \ \sum\nolimits_{k \in {{\cal U}_i}} {{\rm{Tr}}} \left( {{\bf{\bar V}}_k^H{{\bf{B}}_{i,k}}{{{\bf{\bar V}}}_k}} \right) \le {P_{i,\max }},\forall i,
\end{array}
\end{equation}
where  ${c_k} = \log \left| {{{\bf{W}}_k}} \right| + d - {R_{k,\min }} - {\rm{Tr}}\left( {{{\bf{W}}_k}} \right) - \sigma  _k^2{\rm{Tr}}\left( {{\bf{U}}_k^H{{\bf{U}}_k}{{\bf{W}}_k}} \right)$.

\begin{algorithm}
\caption{WMMSE Algorithm}\label{wmmse}
\begin{algorithmic}[1]
\STATE Initialize iterative number $l=1$, maximum number of iterations $l_{\rm{max}}$,  feasible ${\bf{V}}^{(0)}$, calculate ${\bf{U}}^{(0)}$ and ${\bf{W}}^{(0)}$ by using (\ref{receanu}) with ${\bf{V}}^{(0)}$, tolerance $\varepsilon$, calculate the objective value of Problem (\ref{Equivalentpro}), denoted as ${\rm{Obj(}}{{\bf{V}}^{(l - 1)}}{\rm{)}}$.
\STATE With ${\bf{U}}^{(l-1)}$ and ${\bf{W}}^{(l-1)}$, update ${\bf{V}}^{(l)}$ by solving Problem (\ref{equalprosub}) with the BCD algorithm in Subsection \ref{equasolve};
 \STATE Update ${\bf{U}}^{(l)}$ and ${\bf{W}}^{(l)}$ as in (\ref{receanu}) with ${\bf{V}}^{(l)}$;
 \STATE If $l\geq l_{\rm{max}}$ or ${{\left| {{\rm{Obj(}}{{\bf{V}}^{(l - 1)}}{\rm{) - Obj(}}{{\bf{V}}^{(l)}}{\rm{)}}} \right|} \mathord{\left/
 {\vphantom {{\left| {{\rm{Obj(}}{{\bf{V}}^{(l - 1)}}{\rm{) - Obj(}}{{\bf{V}}^{(l)}}{\rm{)}}} \right|} {{\rm{Obj(}}{{\bf{V}}^{(l)}}{\rm{)}}}}} \right.
 \kern-\nulldelimiterspace} {{\rm{Obj(}}{{\bf{V}}^{(l)}}{\rm{)}}}} < \varepsilon $, terminate.  Otherwise, set $l \leftarrow l + 1$  and go to step 2.
\end{algorithmic}
\end{algorithm}

Then, an WMMSE algorithm is proposed to solve Problem (\ref{subproblem}) in Algorithm \ref{wmmse}. In the following theorem, we show  the property of the WMMSE algorithm.

\itshape \textbf{Theorem 2:}  \upshape The sequence of ${\bf{V}}$ generated by the WMMSE algorithm will converge to  the KKT point of Problem (\ref{subproblem}).

\itshape \textbf{Proof:}  \upshape Please see Appendix \ref{prooftheorem2}. \hfill $\Box$

\subsection{Low-complexity Algorithm to Solve Problem (\ref{equalprosub})}\label{equasolve}

 Since $\omega_i>0,\forall i$, matrices $\{{\bf{G}}_k,\forall k\in \cal U\}$ are positive definite matrices. Then,  Problem (\ref{equalprosub}) can be  similarly transformed an SOCP problem as in (\ref{socpfea}). Using the same method in Subsection \ref{comana}, the total complexity to solve this problem by using the interior point method  is
\begin{equation}\label{SOCPcomp}
  {T_{{\rm{SOCP}}}} = O\left( {\sqrt {I + K} {{\left( {2lMKd} \right)}^2}\left( {2{K^2}{d^2} + 2dMKl} \right)} \right).
\end{equation}

In the following, we go one step further to design an algorithm with lower complexity. Obviously,  Problem (\ref{equalprosub}) is a convex problem, and it satisfies the Slater's condition \cite{boyd2004convex}. Hence, the duality gap between Problem (\ref{equalprosub}) and its dual problem is zero \cite{boyd2004convex}.  Then we can solve its dual problem instead of directly solving it.

With some simple manipulations, the Lagrangian function of Problem (\ref{equalprosub}) is given by
\begin{eqnarray}
{\cal L}\left( {{\bf{V}},{\bm{\lambda}},{\bm{\mu}} } \right)& = & \sum\nolimits_{k \in {\cal U}} {\left( {{\rm{Tr}}\left( {{\bf{\bar V}}_k^H{{{\bf{\bar G}}}_k}{{{\bf{\bar V}}}_k}} \right) - {\rm{Tr}}\left( {{\lambda _k}{{\bf{W}}_k}{\bf{U}}_k^H{{{\bf{\bar H}}}_{k,k}}{{{\bf{\bar V}}}_k}} \right) - {\rm{Tr}}\left( {{\lambda _k}{\bf{\bar V}}_k^H{\bf{\bar H}}_{k,k}^H{{\bf{U}}_k}{{\bf{W}}_k}} \right)} \right)} \nonumber\\
 && - \sum\nolimits_{k \in {\cal U}} {{\lambda _k}{c_k}}  - \sum\nolimits_{i \in {\cal I}} {{\mu _i}{P_{i,\max }}}, \nonumber
\end{eqnarray}
where ${\bm{\lambda}} = [{{\lambda _k}, \forall k\in \cal U} ]^H$ and ${\bm{\mu}}  = [ {{\mu _i}, \forall i\in \cal I} ]^H$  are  the Lagrangian multipliers associated with the first and second sets of constrains of Problem (\ref{equalprosub}), respectively,  and  ${\bf{\bar G}}_k$ is given by
 \[{{\bf{\bar G}}_k} = {{\bf{G}}}_k + \sum\nolimits_{j \in \cal U} {{\lambda _j}{\bf{\bar H}}_{k,j}^H{{\bf{U}}_j}{{\bf{W}}_j}{\bf{U}}_j^H{{\bf{\bar H}}_{k,j}}}  + \sum\nolimits_{i\in {\cal I}_k} {{\mu _i}} {{\bf{B}}_{i,k}}.\]
 The dual function is given by

\begin{eqnarray}
g({\bm{\lambda}},{\bm{\mu}} )&=& \mathop {{\rm{min}}}\limits_{\bf{V}} {\cal L} \left( {{\bf{V}},{\bm{\lambda}},{\bm{\mu}}} \right)\nonumber\\
 &=& \mathop {{\rm{min}}}\limits_{\bf{V}}  \sum\nolimits_{k \in {\cal U}} {\left( {{\rm{Tr}}\left( {{\bf{\bar V}}_k^H{{{\bf{\bar G}}}_k}{{{\bf{\bar V}}}_k}} \right) - {\rm{Tr}}\left( {{\lambda _k}{{\bf{W}}_k}{\bf{U}}_k^H{{{\bf{\bar H}}}_{k,k}}{{{\bf{\bar V}}}_k}} \right) - {\rm{Tr}}\left( {{\lambda _k}{\bf{\bar V}}_k^H{\bf{\bar H}}_{k,k}^H{{\bf{U}}_k}{{\bf{W}}_k}} \right)} \right)} \nonumber\\
 &&\qquad - \sum\nolimits_{k \in {\cal U}} {{\lambda _k}{c_k}}  - \sum\nolimits_{i \in {\cal I}} {{\mu _i}{P_{i,\max }}}.\label{dualfunctionprob}
\end{eqnarray}
Note that matrices $\{{\bf{G}}_k,\forall k\in \cal U\}$ are positive definite matrices. Problem (\ref{dualfunctionprob}) is a convex problem, and the optimal solution can be obtained from its first-order derivative condition as:
\begin{equation}\label{optimalv}
{{{\bf{\bar V}}}_k} = {\lambda _k}{\bf{\bar G}}_k^{ - 1}{\bf{\bar H}}_{k,k}^H{{\bf{U}}_k}{{\bf{W}}_k},\forall k \in {\cal U}.
\end{equation}

By inserting this solution into (\ref{dualfunctionprob}), the dual function becomes
\begin{equation}\label{dualfunction}
 g({\bm{\lambda}},{\bm{\mu}} ) =  - \sum\nolimits_{k \in {\cal U}} {\lambda _k^2{\rm{Tr}}\left( {{\bf{W}}_k^H{\bf{U}}_k^H{{{\bf{\bar H}}}_{k,k}}{\bf{\bar G}}_k^{ - 1}{\bf{\bar H}}_{k,k}^H{{\bf{U}}_k}{{\bf{W}}_k}} \right)}   - \sum\nolimits_{k \in {\cal U}} {{\lambda _k}{c_k}}  - \sum\nolimits_{i \in {\cal I}} {{\mu _i}{P_{i,\max }}}.
\end{equation}

Hence, the dual problem of Problem (\ref{equalprosub}) is given by
\begin{eqnarray}
&&\mathop {\max }\limits_{\{ {\lambda _k}\geq 0,\forall k\},\{ {\mu _i}\geq 0,\forall i\}} g({\bm{\lambda}},{\bm{\mu}} )\nonumber\\
 &=&\!\! \mathop {\min }\limits_{\{ {\lambda _k}\geq 0,\forall k\},\{ {\mu _i}\geq 0,\forall i\}}\! \sum\nolimits_{k \in {\cal U}} {\lambda _k^2{\rm{Tr}}\left( {{\bf{W}}_k^H{\bf{U}}_k^H{{{\bf{\bar H}}}_{k,k}}{\bf{\bar G}}_k^{ - 1}{\bf{\bar H}}_{k,k}^H{{\bf{U}}_k}{{\bf{W}}_k}} \right)} \!+\! \sum\nolimits_{k \in {\cal U}} {{\lambda _k}{c_k}} \! +\! \sum\nolimits_{i \in {\cal I}} {{\mu _i}{P_{i,\max }}} \nonumber\\
 &\triangleq&  \mathop {\min }\limits_{\{ {\lambda _k}\geq 0,\forall k\},\{ {\mu _i}\geq 0,\forall i\}} f({\bm{\lambda}},{\bm{\mu}} ), \label{dualproblem}
\end{eqnarray}
where $f({\bm{\lambda}},{\bm{\mu}} ) =  - g({\bm{\lambda}},{\bm{\mu}})$.

Fortunately, the objective function of the dual problem in (\ref{dualproblem}) is differentiable and the dual problem is convex \cite{boyd2004convex}, the descent methods such as the gradient descent  method and Newton's  method \cite{boyd2004convex,bertsekas1999nonlinear} can be applied to solve it. In the following, we also utilize the block coordinate descent method to solve the dual problem (\ref{dualproblem}): Optimize $\left\{ {{\lambda _k}},\forall k \right\}$ with  $\left\{ {{\mu _i}},\forall i \right\}$, and vice versa.

Given $\left\{ {{\mu _i}},\forall i \right\}$,  Newton's method is applied to find the optimal $\left\{ {{\lambda _k}},\forall k \right\}$  of the dual problem, which is summarized in Algorithm \ref{newtons}. \footnote{Since $\left\{ {{\mu _i}},\forall i \right\}$ are given, $f({\bm{\lambda}})$ is short for $f({\bm{\lambda}},{\bm{\mu}} )$ and the same for $f({\bm{\mu}})$ later.}

\begin{algorithm}
\caption{Newton's Method to Update $\left\{ {\lambda _k},\forall k \right\}$}\label{newtons}
\begin{algorithmic}[1]
\STATE Initialize iterative number $t=1$, the maximum number of iterations $t_{\rm{max}}^{\rm{Newt}}$, initial  $\bm{\lambda}^{(0)}=\bf{1} $, tolerance $\varepsilon  = {10^{ - 10}}$,  $\xi  \in (0,0.5)$, $\varphi  \in (0,1)$;
\STATE Compute the gradient $\nabla f(\bm{\lambda}^{(t-1)} )$,  Hessian matrix $\nabla^2 f(\bm{\lambda}^{(t-1)} )$, the Newton direction and the decrement
\[\Delta {\bm{\lambda}^{(t-1)}}  =  - {\left( {{\nabla ^2}f({\bm{\lambda}^{(t-1)}}  )} \right)^{ - 1}}\nabla f({\bm{\lambda}^{(t-1)}}), o^{(t-1)} = \nabla f{({\bm{\lambda}^{(t-1)}})^T}{\left( {{\nabla ^2}f({\bm{\lambda}^{(t-1)}} )} \right)^{ - 1}}\nabla f({\bm{\lambda}^{(t-1)}} );\]
 \STATE Compute  $\bm{\bar \lambda}^{(t-1)}  = {[\bm{\lambda}^{(t-1)}  + \Delta {\bm{\lambda}}^{(t-1)} ]_ + }$;
 \STATE Update  ${\bm{\lambda}}^{(t )} = {\bm{\lambda}}^{(t-1)} + {\kappa ^{(t-1)}}( \bm{\bar \lambda}^{(t-1)} - {\bm{\lambda}}^{(t-1)} )$, where ${\kappa  ^{(t-1)}} = {\varphi  ^{{m^{(t-1)}}}}$  and  ${m^{(t-1)}}$ is the first non-negative integer $m$  that satisfies
 \begin{equation}\label{stopcondition}
   f({{\bm{\lambda}} ^{(t)}}) - f({{\bm{\lambda}} ^{(t-1)}}) \le \xi  {\varphi  ^m}\nabla f{({{\bm{\lambda}} ^{(t-1)}})^T}\left( {{{\bar {\bm{\lambda}} }^{(t-1)}} - {{\bm{\lambda}} ^{(t-1)}}} \right).
 \end{equation}
\STATE If $o^{(t-1)}/2\leq \varepsilon$ or $t \ge {t_{\max }^{\rm{Newt}}}$, terminate; Otherwise, $t \leftarrow t + 1$, and go to step 2;
\end{algorithmic}
\end{algorithm}

In step 4 of Algorithm \ref{newtons}, the backtracking line search method is used to find the step size, where $\xi $ is typically chosen as a very small value and $\varphi $ is chosen between $0$ and $1$. The step $\kappa ^{(t)}$ starts with one and then reduces by a factor of $\varphi $ until the stop condition (\ref{stopcondition}) is satisfied. Note that in each iteration of Algorithm \ref{newtons}, the step value $\kappa ^{(t)}$ may be different. The constant $\xi $ can be regarded as the acceptable fraction of the decrease in the objective value of $f$  that is predicted by the line search method.

However, to make this algorithm work, there are still problems to be solved: how to calculate the gradient and how to compute the Hessian matrix. To derive the expressions of the gradient and the Hessian matrix, we first introduce some useful results in the matrix differential calculus. Given a matrix function $\bm{\Gamma}(x)$, one has \cite{magnus1995matrix,Sigen2003}
\begin{equation}\label{derivativetrace}
 \frac{d}{{dx}}{\rm{Tr}}\left( \bm{\Gamma}(x) \right) = {\rm{Tr}}\left( {\frac{{d\bm{\Gamma}(x)}}{{dx}}} \right),
\end{equation}
\begin{equation}\label{derivativeinv}
\frac{d}{{dx}}\bm{\Gamma} {(x)^{ - 1}} =  - \bm{\Gamma} {(x)^{ - 1}}\frac{{d\bm{\Gamma} (x)}}{{dx}}\bm{\Gamma} {(x)^{ - 1}}.
\end{equation}

 In addition, to simplify the expressions of  the gradient and the Hessian matrix, one defines some matrices:
 \begin{equation}\label{singlematrix}
   \begin{array}{l}
{{{\bf{\tilde H}}}_{j,k}} = {\bf{\bar H}}_{j,k}^H{{\bf{U}}_k},{{{\bf{\mathord{\buildrel{\lower3pt\hbox{$\scriptscriptstyle\smile$}}
\over H} }}}_{j,k}} = {{{\bf{\tilde H}}}_{j,k}}{{\bf{W}}_k},{{{\bf{\hat H}}}_{j,k}} = {{{\bf{\mathord{\buildrel{\lower3pt\hbox{$\scriptscriptstyle\smile$}}
\over H} }}}_{j,k}}{\bf{\tilde H}}_{j,k}^H,{{{\bf{\tilde G}}}_k} = {\bf{\bar G}}_k^{ - 1},{{\bf{C}}_k} = {{{\bf{\tilde G}}}_k}{{{\bf{\mathord{\buildrel{\lower3pt\hbox{$\scriptscriptstyle\smile$}}
\over H} }}}_{k,k}},\\
{{\bf{F}}_k} = {\bf{\mathord{\buildrel{\lower3pt\hbox{$\scriptscriptstyle\smile$}}
\over H} }}_{k,k}^H{{\bf{C}}_k},{{\bf{Y}}_{j,k}} = {\bf{C}}_j^H{{{\bf{\hat H}}}_{j,k}},{{{\bf{\tilde Y}}}_{j,k}} = {{\bf{Y}}_{j,k}}{{{\bf{\tilde G}}}_j},{{\bf{Z}}_{j,k}} = {{\bf{Y}}_{j,k}}{{\bf{C}}_j},\forall j,k \in \cal U.
\end{array}
 \end{equation}

Based on the above results and definitions, the gradient can be derived as follows:
\begin{equation}\label{gradient}
\nabla f(\bm{\lambda} ) = {\left[ {\frac{{\partial f(\bm{\lambda} )}}{{\partial{\lambda _k}}}, \forall k\in \cal U} \right]^H},
\end{equation}
with
\begin{equation}\label{firstdire}
  \frac{{\partial f(\bm{\lambda} )}}{{\partial {\lambda _k}}} = 2{\lambda _k}{\rm{Tr}}\left( {{{\bf{F}}_k}} \right) - \sum\nolimits_{j\in \cal U} {\lambda _j^2{\rm{Tr}}} \left( {{{\bf{Z}}_{j,k}}} \right) + {c_k},k \in \cal U.\vspace{-0.2cm}
\end{equation}
The Hessian matrix of $f(\bm{\lambda})$ can be calculated as:
\begin{equation}\label{hessian}
{\left[ {{\nabla ^2}f({\bm{\lambda}} )} \right]_{i,j}} = \left\{ \begin{array}{l}
2{\rm{Tr}}({{\bf{F}}_i}) + 2\sum\nolimits_{k \in {\cal U}} {\lambda _k^2{\rm{Tr}}\left( {{{{\bf{\tilde Y}}}_{k,i}}{\bf{Y}}_{k,i}^H} \right)}  - 4{\lambda _i}{\rm{Tr}}\left( {{{\bf{Z}}_{i,i}}} \right),\qquad\qquad\ \   \ {\rm{ if  }}\ i = j,\\
- 2{\lambda _i}{\rm{Tr}}\left( {{{\bf{Z}}_{i,j}}} \right) - 2{\lambda _j}{\rm{Tr}}\left( {{{\bf{Z}}_{j,i}}} \right) + 2\sum\nolimits_{k \in {\cal U}} {\lambda _k^2{\mathop{\rm Re}\nolimits} \left\{ {{\rm{Tr}}\left( {{{{\bf{\tilde Y}}}_{k,j}}{\bf{Y}}_{k,i}^H} \right)} \right\}} {\rm{,if}}\;j > i,\\
{\left[ {{\nabla ^2}f({\bm{\lambda}} )} \right]_{j,i}},\qquad\qquad\qquad\qquad\qquad\qquad\qquad\qquad\quad\  \qquad\ \quad\ {\rm{  if }}\ j < i.
\end{array} \right.
\end{equation}

Next,  given $\left\{ {{\lambda _k}}, \forall k\in \cal U \right\}$, we solve the dual problem (\ref{dualproblem}) to update $\left\{ {{\mu _i}},\forall i\in \cal I \right\}$. Here, the gradient descent method \cite{boyd2004convex} is applied. Although  Newton's method converges faster than the gradient descent method, simulation results show that the gradient method also converges within five iterations but it has much lower computational complexity than  Newton's method since it does not require the calculations of the Hessian matrix and  the inverse of the Hessian matrix. The gradient descent method to update $\left\{ {{\mu _i}},\forall i\in\cal I \right\}$ is given in  Algorithm \ref{gradient}.

In Algorithm  \ref{gradient}, the gradient $\nabla f(\bm{\mu } )$ is required. Define ${{\bf{D}}_k} = {{\bf{C}}_k}{\bf{C}}_k^{\rm{H}}$. Then, by using the results in (\ref{derivativetrace}) and (\ref{derivativeinv}), the gradient $\nabla f(\bm{\mu } )$ can be calculated as
\begin{equation}\label{gradientmu}
\nabla f(\bm{\mu} ) = {\left[ {\frac{{df(\bm{\mu} )}}{{d{\mu _i}}}, \forall i\in\cal I} \right]^H},\vspace{-0.2cm}
\end{equation}
with
\begin{equation}\label{firstdiremu}
  \frac{{df(\bm{\mu} )}}{{d{\mu _i}}} =  - \sum\nolimits_{k \in {{\cal U}_i}} {\lambda _k^2{\rm{Tr}}\left( {{{\bf{B}}_{i,k}}{{\bf{D}}_k}} \right)}  + {P_{i,\max }},\forall i \in {\cal I}.
\end{equation}

\begin{algorithm}
\caption{Gradient Descent Method to Update $\left\{ {{\mu _i}} \right\}_{i = 1}^I$}\label{gradient}
\begin{algorithmic}[1]
\STATE Initialize iterative number $t=1$, maximum number of iterations $t_{\rm{max}}^{\rm{Grad}}$, initial  $\bm{\mu}^{(0)}=\bf{1} $, accuracy $\varepsilon$;
\STATE Compute the gradient $\nabla f(\bm{\mu }^{(t-1)} )$;
 \STATE Compute  $\bm{\bar \mu}^{(t-1)}  = {[\bm{\mu}^{(t-1)}  -  \nabla f(\bm{\mu }^{(t-1)} ) ]_ + }$;
 \STATE Update  ${\bm{\mu}}^{(t )} = {\bm{\mu}}^{(t-1)} + {\kappa ^{(t-1)}}( \bm{\bar \mu} - {\bm{\mu}}^{(t-1 )} )$, where ${\kappa  ^{(t-1)}} = {\beta ^{{l^{(t-1)}}}}$  and  ${l^{(t-1)}}$ is the first non-negative integer $l$  that satisfies
\[f({{\bm{\mu}} ^{(t )}}) - f({{\bm{\mu}} ^{(t-1)}}) \le \delta {\beta ^l}\nabla f{({{\bm{\mu}} ^{(t-1)}})^T}\left( {{{\bar {\bm{\mu}} }^{(t-1)}} - {{\bm{\mu}} ^{(t-1)}}} \right). \]
\STATE If $t \ge {t_{\max }}$ or ${{\left| {f({\bm{\mu} ^{(t)}}) - f({\bm{\mu} ^{(t - 1)}})} \right|} \mathord{\left/
 {\vphantom {{\left| {f({\mu ^{(t)}}) - f({\mu ^{(t - 1)}})} \right|} {\left| {f({\mu ^{(t)}})} \right|}}} \right.
 \kern-\nulldelimiterspace} {\left| {f({\bm{\mu} ^{(t)}})} \right|}} < \varepsilon $, stop; Otherwise, $t \leftarrow t + 1$, and go to step 2;
\end{algorithmic}
\end{algorithm}

Finally, based on the above analysis, the  method to solve the dual problem (\ref{dualproblem}) is given in Algorithm \ref{bcdd}, which is named as Block Coordinate Descent  (BCD) method.
\begin{algorithm}
\caption{BCD  Method to Solve the Dual Problem (\ref{dualproblem})}\label{bcdd}
\begin{algorithmic}[1]
\STATE Initialize iterative number $n=1$, the maximum number of iterations $n_{\rm{max}}$, initial  $\bm{\lambda}^{(0)}=\bf{1} $ and $\bm{\mu}^{(0)}=\bf{1}$, error tolerance $\varepsilon$;
\STATE Given $\bm{\mu}^{(n-1)}$, apply Newton's method in Algorithm \ref{newtons} to update $\bm{\lambda}^{(n)}$;
\STATE Given $\bm{\lambda}^{(n)}$, employ the gradient descent method in Algorithm \ref{gradient} to update $\bm{\mu}^{(n)}$;
\STATE If $n \ge {n_{\max }}$ or ${{\left| {f({\bm{\lambda} ^{(n)}},{\bm{\mu} ^{(n)}}) - f({\bm{\lambda} ^{(n-1)}},{\bm{\mu} ^{(n - 1)}})} \right|} \mathord{\left/
 {\vphantom {{\left| {f({\lambda ^{(n)}},{\mu ^{(n)}}) - f({\lambda ^{(t)}},{\mu ^{(n - 1)}})} \right|} {\left| {f({\lambda ^{(n)}},{\mu ^{(n)}})} \right|}}} \right.
 \kern-\nulldelimiterspace} {\left| {f({\bm{\lambda} ^{(n)}},{\bm{\mu} ^{(n)}})} \right|}} < \varepsilon $, terminate; Otherwise, set $n \leftarrow n + 1$, and go to step 2;
\end{algorithmic}
\end{algorithm}

\itshape \textbf{Theorem 3:}  \upshape The sequences of $\bm{\mu}$ and $\bm{\lambda}$ generated by Algorithm \ref{bcdd} will converge to the globally optimal solution of the dual problem (\ref{dualproblem}).

\itshape \textbf{Proof:}  \upshape Please see Appendix \ref{prooftheorem3}. \hfill $\Box$

When the optimal $\bm{\mu}$ and $\bm{\lambda}$ are obtained by using Algorithm \ref{bcdd}, the optimal solution to Problem (\ref{equalprosub}) is given by (\ref{optimalv}). As there is zero duality gap between the primal problem (\ref{equalprosub}) and dual  problem (\ref{dualproblem}), which means that this solution   is the globally optimal solution of Problem (\ref{equalprosub}).

\subsection{Overall Complexity to Solve Problem (\ref{equalprosub}) in Stage II}
In this subsection, we analyze the overall complexity to solve Problem (\ref{equalprosub}). It mainly includes three layers of iterations: the first layer is the RLN algorithm to deal with the non-smooth $l_0$ norm, the second layer is the WMMSE algorithm to deal with the non-convex rate constraints, and the third layer is the BCD algorithm to solve Problem (\ref{equalprosub}).

We first analyze the complexity of the third layer for BCD algorithm. Note that ${{{\bf{\tilde H}}}_{j,k}}$, ${{{\bf{\mathord{\buildrel{\lower3pt\hbox{$\scriptscriptstyle\smile$}}\over H} }}}_{j,k}}$, and ${{{\bf{\hat H}}}_{j,k}} $ can be calculated before the iterations of the BCD Algorithm. The main complexity of the BCD Algorithm lies in step 2 and step 3, where  Newton's method and gradient descent method are used to update ${\bm{\lambda}}$ and $\bm{\mu}$, respectively.

We first analyze the computational complexity of Newton's method under the same assumption in Subsection \ref{comana}. The main complexity in each iteration of Newton's method lies in step 2 and step 4 of Algorithm \ref{newtons}. We first analyze step 2 of Algorithm \ref{newtons}. According to \cite{coppersmith1987matrix}, the complexity of calculating $\{{{\bf{\tilde G}}}_k,\forall k\in\cal U\}$ is on the order of $O\left( {{{K\left( {Ml} \right)}^{2.376}}} \right)$. For any two matrices ${\bf{X}} \in {{\mathbb{C}}^{m \times n}},{\bf{Y}} \in {{\mathbb{C}}^{n \times p}}$, the complexity of computing ${\bf{XY}}$ is on the order of $O\left( {mnp} \right)$ \cite{boyd2004convex}. In general, $d \ll MI$. Then, the total complexity of computing $\left\{ {{{\bf{C}}_k},{{\bf{F}}_k},\forall k\in\cal U} \right\}$ is on the order of $O\left( {K{M^2}{l^2}d} \right)$. Similarly, the total complexity of computing $\left\{ {{{\bf{Y}}_{j,k}},{{{\bf{\tilde Y}}}_{j,k}},{{\bf{Z}}_{j,k}}}\forall j,k\in\cal U \right\}$ is on the order of $O\left( {{K^2}{M^2}{l^2}d} \right)$. Hence, the total complexity of computing $\left\{ {{{\bf{C}}_k},{{\bf{F}}_k}},\forall k \right\}$ and $\left\{ {{{\bf{Y}}_{j,k}},{{{\bf{\tilde Y}}}_{j,k}},{{\bf{Z}}_{j,k}},\forall j,k\in\cal U} \right\}$ is on the order of $O\left( {{K^2}{M^2}{l^2}d} \right)$. With a similar analysis, the total complexity of computing (\ref{hessian})  is on the order of $O\left( {{K^3}Ml{d^2}} \right)$. In addition, the complexity of computing the inverse of $\nabla^2 f(\bm{\lambda}^{(t)} )$ is on the order of $O\left( {{K^{2.376}}} \right)$ \cite{coppersmith1987matrix}.
 Hence, the total complexity of step 2 of Algorithm \ref{newtons} is on the order of $O\left( {\max \left\{ {{K^3}Ml{d^2},{{\left( K{Ml} \right)}^{2.376}},{K^2}{{\left( {Ml} \right)}^2}d} \right\}} \right)$. In the $t$th iteration of  step 4 of Algorithm \ref{newtons}, $f({{\bm{\lambda}} ^{(t + 1)}})$ is required to calculate $m^{(t)}$ times. The complexity in each time is on the order of $O\left( {\max \left\{ {{{\left( {Ml} \right)}^{2.376}},K{{\left( {Ml} \right)}^2}d} \right\}} \right)$. Thus, the total complexity of step 4 of Algorithm \ref{newtons}  is on the order of $O\left({m^{(t)}} {\max \left\{ {{{\left( {Ml} \right)}^{2.376}},K{{\left( {Ml} \right)}^2}d} \right\}} \right)$. Simulation results show that in general $m^{(t)}$ is always equal to one, which means that $f({{\bm{\lambda}} ^{(t + 1)}})$  only needs to be computed for  once. Hence, the complexity of step 4 of Algorithm \ref{newtons} can be approximately by  $O\left({\max \left\{ {{{\left( {Ml} \right)}^{2.376}},K{{\left( {Ml} \right)}^2}d} \right\}} \right)$. As a result, the total complexity of  Newton's method is
 \vspace{-0.2cm}
 \begin{equation}\label{newtoncom}
 T_{\rm{Newton}}= O\left( t_{\rm{max}}^{\rm{Newt}}{\max \left\{ {{K^3}Ml{d^2},{{\left( {Ml} \right)}^{2.376}},{K^2}{{\left( {Ml} \right)}^2}d} \right\}} \right).\vspace{-0.2cm}
 \end{equation}
Simulation results show that Newton's method converges very rapidly and in general five iterations are enough for the algorithm to converge.

By using the similarly analytical technique to Newton's method, the total complexity of the gradient descent method is given by
\begin{equation}\label{complexity}
  T_{\rm{Grad}}=O\left( {t_{{\rm{max}}}^{{\rm{Grad}}}\max \left\{ {{{\left( {MI} \right)}^{2.376}},K{{\left( {MI} \right)}^2}d} \right\}} \right).
\end{equation}
The simulation results in the next section show that the gradient descent method usually converges within five iterations. Hence, in each iteration of the BCD Algorithm, the complexity of Newton's method dominates the complexity of the gradient descent method.

Based on the above analysis, the overall complexity to solve Problem (\ref{equalprosub}) in Stage II is
 \begin{equation}\label{complexitystageII}
   {T_{{\rm{StageII}}}} = {t_{{\rm{RLN}}}}{t_{{\rm{WMMSE}}}}{t_{{\rm{BCD}}}}\left( {{T_{{\rm{Newton}}}} + {T_{{\rm{Grad}}}}} \right),
 \end{equation}
where $t_{{\rm{RLN}}}$, $t_{{\rm{WMMSE}}}$ and  $t_{{\rm{BCD}}}$ represent the average number of iterations required by the RLN, WMMSE, and BCD algorithms, respectively. Simulation results show that these three algorithms converge very fast and generally five iterations are enough to achieve large portion of the final performance.

\section{Simulation Results}\label{simulation}

\begin{figure}
\begin{minipage}[t]{0.495\linewidth}
\centering
\includegraphics[width=2.2in]{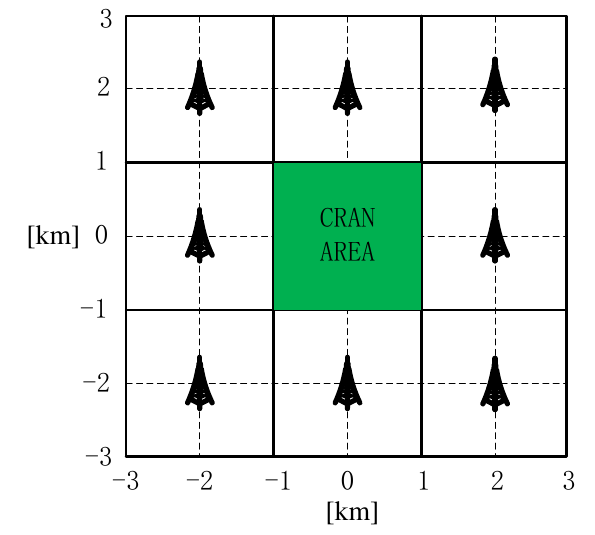}
\caption{Illustration of a wrap-round C-RAN system model, where C-RAN is deployed in the center of the region, which is surrounded by eight nearby cells. }
\label{simulationmodel}
\end{minipage}%
\hfill
\begin{minipage}[t]{0.495\linewidth}
\centering
\includegraphics[width=2.6in]{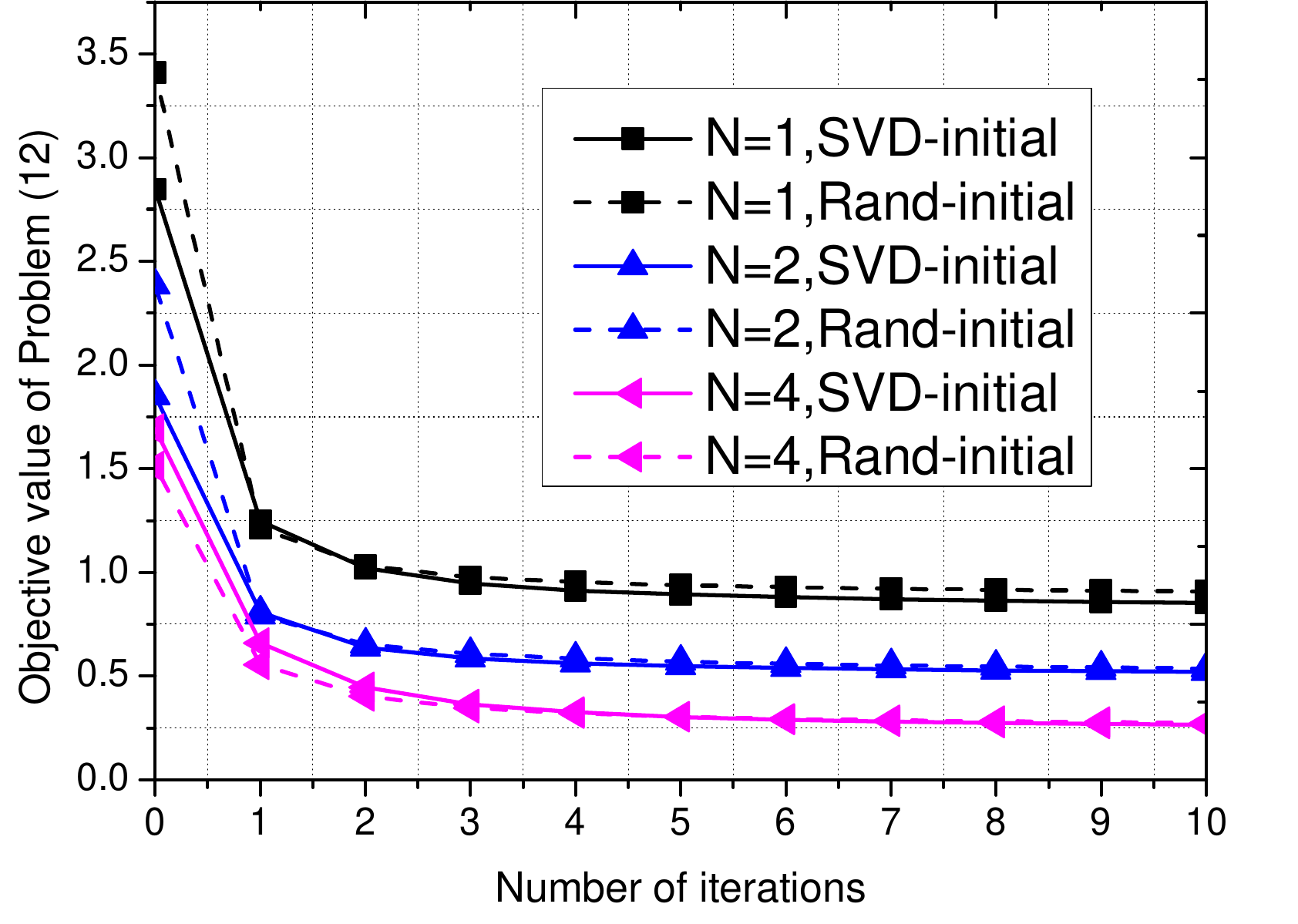}
\caption{Convergence behaviour of Algorithm 2.}
\label{Algocon2}
\end{minipage}
\end{figure}

In this section, we present simulation results to evaluate the performance of the proposed algorithms. To be more realistic, we consider a wrap-around system model shown in Fig. \ref{simulationmodel} as in \cite{Venturino2010}, where the C-RAN network is deployed in the central square with $[ - 1000\ {\rm{ 1000}}] \times [ - 1000\ {\rm{ 1000}}]$ meters, surrounded by eight uncoordinated square macrocells. It is assumed that all the users and RRHs are uniformly and independently distributed in the C-RAN region. We adopt the channel model that consists of four parts: 1) the  long term evolution (LTE) standard path loss model: $PL_{i,k} = 148.1 + 37.6{\log _{10}}d_{i,k}\ ({\rm{dB}})$, where $d_{i,k}$ (in km) is the distance from the $i$th RRH to the $k$th user; 2) Log-normal shadowing with zero mean and 8 dB standard derivation; 3) Rayleigh fading with zero mean and unit variance; 4) transmit antenna power gain of 9 dBi.  Each user is assumed to have the same rate requirement, i.e., $R_{\rm{min}}=R_{k,\rm{min}}, \forall k$, and each RRH has the same power constraint, i.e., $P_{\rm{max}}=P_{i,\rm{max}}=4 {\rm{W}}, \forall i\in \cal I$. It is assumed that each user is potentially served by its nearest $X$ RRHs, i.e., $|{{\cal I}_k}|=X, \forall i$.  Unless stated otherwise, the system parameters are set as follows: error tolerance is $\varepsilon=10^{-3}$, thermal noise power is $\sigma^2=-104\  {\rm{dBm}}$, $I=12$, $K=8$, $X=3$, $M=2$, $N=2$,  $d={\rm{min}} \{M,N\}$, $\eta_{i}=4$ \cite{Auer2011}, $\rho_i=0.5$ \cite{dai2016energy},  $P_i^{a,{\rm{rrh}}}=3.4{\rm{W}}$, $P_i^{s,{\rm{rrh}}}=2.15{\rm{W}}$, $P_i^{a,{\rm{fr}}}=3.85 {\rm{W}}$, $P_i^{s,{\rm{fr}}}=0.75 {\rm{W}}$, ${P_{{\rm{BBU}}}}=20 W$ \cite{Dhaini,Yuanming2014}. Moreover, let $\cal L$ be the set of uncoordinated base stations (BSs) in C-RAN's nearby eight macrocells. The noise power at user $k$ can be modeled as $\sigma _k^2 = {\sigma ^2} + \sum\nolimits_{m \in {\cal L}} {{P_{\max }}P{L_{m,k}}{S_{m,k}}{G_m}}$ \cite{Venturino2010}, where $P{L_{m,k}}$ and $S_{m,k}$  are the large-scale fading and shadowing  respectively from the BS in macrocell $m$ to user $k$, ${G_m}$ represents the antenna gain.

\subsection{Properties of the Proposed Algorithms}

\subsubsection{Convergence behavior of Algorithm \ref{iterdada}} Fig.~\ref{Algocon2} shows the convergence behaviour of Algorithm \ref{iterdada} for different numbers of receive antennas. The results are obtained by averaging over 100 channel realizations. Due to the non-convexity of Problem (\ref{alternativepro}), different initial points for Algorithm \ref{iterdada} may yield different solutions. To investigate this effect, we consider two initialization schemes: 1) SVD-initial, in which the beam directions for each user are chosen as the unitary matrices obtained by the singular value decomposition (SVD) of channel matrices and the total power at each RRH is equally allocated to the users potentially served by each RRH; 2) Rand-initial, in which both the beam directions and power allocations are randomly generated. It can be seen from Fig.~\ref{Algocon2} that the objective value of Problem (\ref{alternativepro}) monotonically decreases during the iterative procedure for two initialization schemes. In addition, the algorithm converges very fast and in general six iterations are sufficient to achieve a large proportion of the converged value for different numbers of receive antennas and different initialization schemes. It is interesting to find that the algorithm under two different initialization schemes will converge to almost the same value. As expected, the converged objective value decreases with the number of receive antennas since more degrees of freedom are available.
\begin{figure}
\begin{minipage}[t]{0.495\linewidth}
\centering
\includegraphics[width=2.6in]{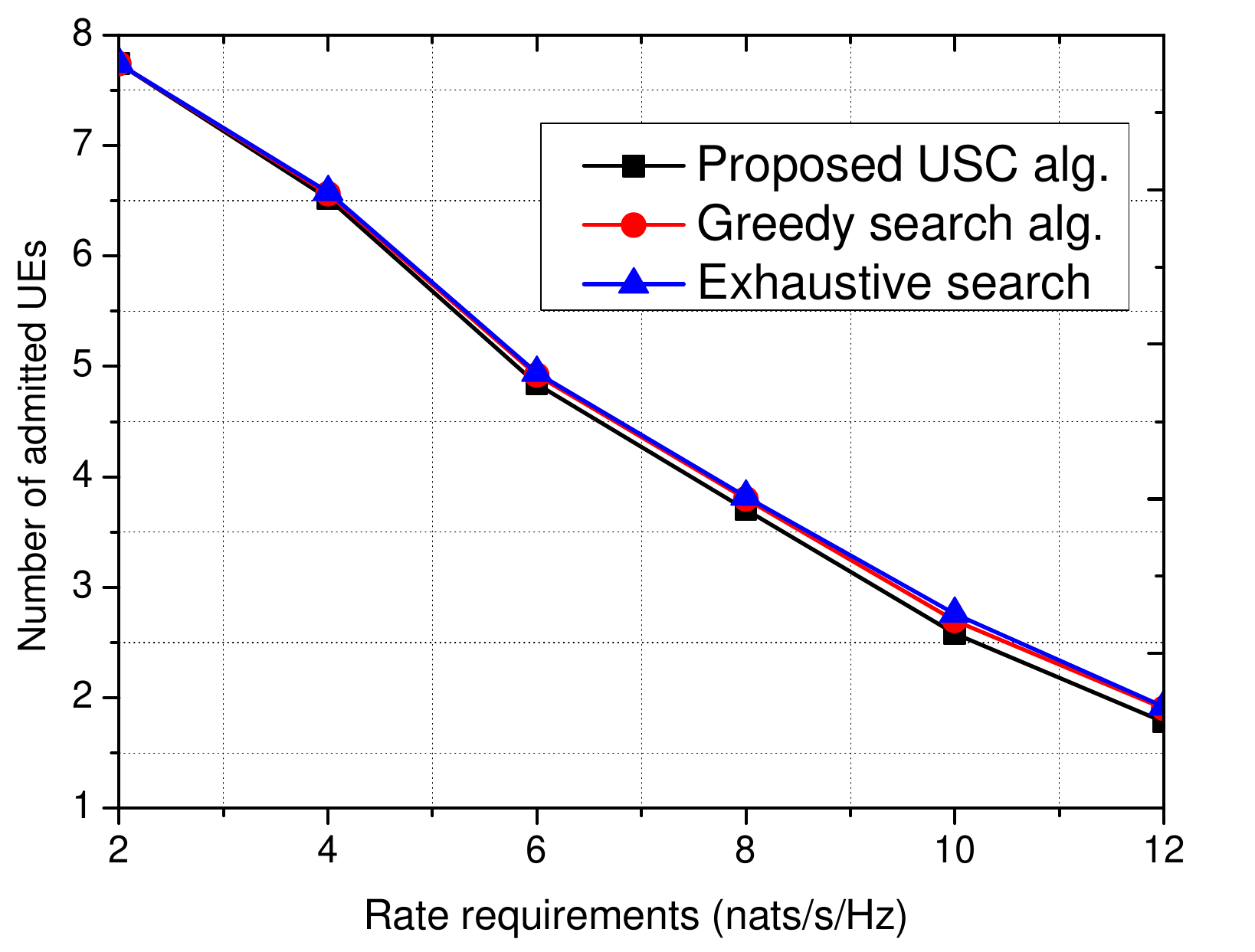}
\caption{Average number of admitted users versus the rate requirements.}
\label{USCper}
\end{minipage}%
\hfill
\begin{minipage}[t]{0.495\linewidth}
\centering
\includegraphics[width=2.9in]{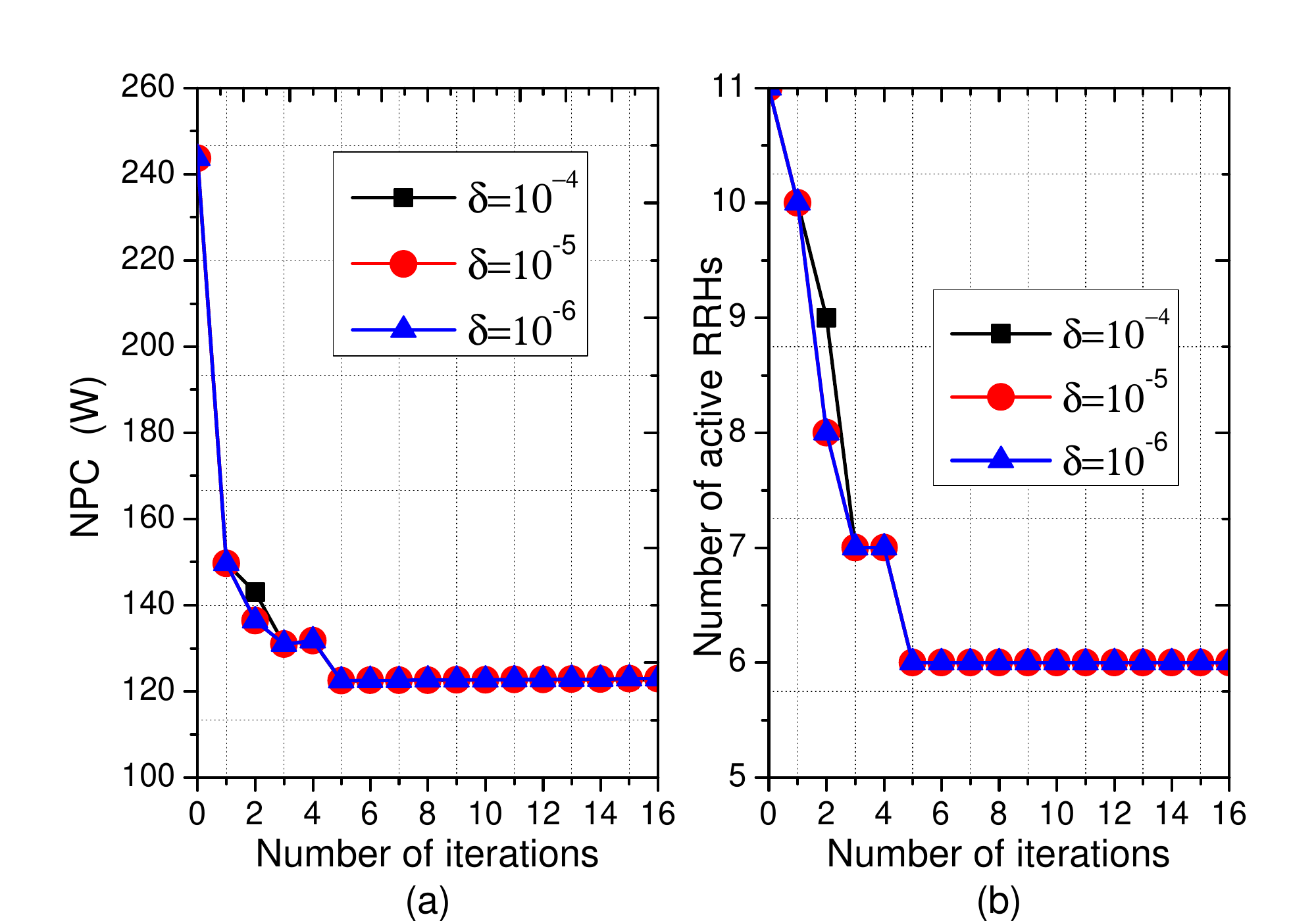}
\caption{(a) Total power consumption versus the number of iterations; (b) The number of active RRHs versus the number of iterations, where $R_{\rm{min}}=2$ nats/s/Hz.}
\label{RLNalgcon}
\end{minipage}
\end{figure}
\subsubsection{User selection performance of USC algorithm}
Fig.~\ref{USCper} compares the performance of the USC algorithm with two algorithms: greedy search method and exhaustive search method. For the greedy search method, in each time we compute the objective value of  Problem (\ref{alternativepro}) when excluding one user, then the user yielding the smallest objective value will be removed. This procedure continues until all remaining users are feasible. Note that this algorithm increases quadratically with $K$. The exhaustive search method checks all feasible sets of users and chooses the largest one. Its complexity increases exponentially with $K$. As expected, the number of admitted users decreases with the rate requirements for all algorithms. The greedy search method achieves almost the same performance as the exhaustive one, and the performance gap between the exhaustive search algorithm and the proposed USC algorithm can be negligible. However, the complexity of our proposed USC algorithm only increases linearly with $K$. The impact of initial points is also studied and  we find that both initialization schemes (SVD-initial and rand-initial) have similar performance, which is not shown here for clarity.

\subsubsection{Convergence behaviour of the RLN algorithm}

\begin{figure}
\begin{minipage}[t]{0.495\linewidth}
\centering
\includegraphics[width=2.6in]{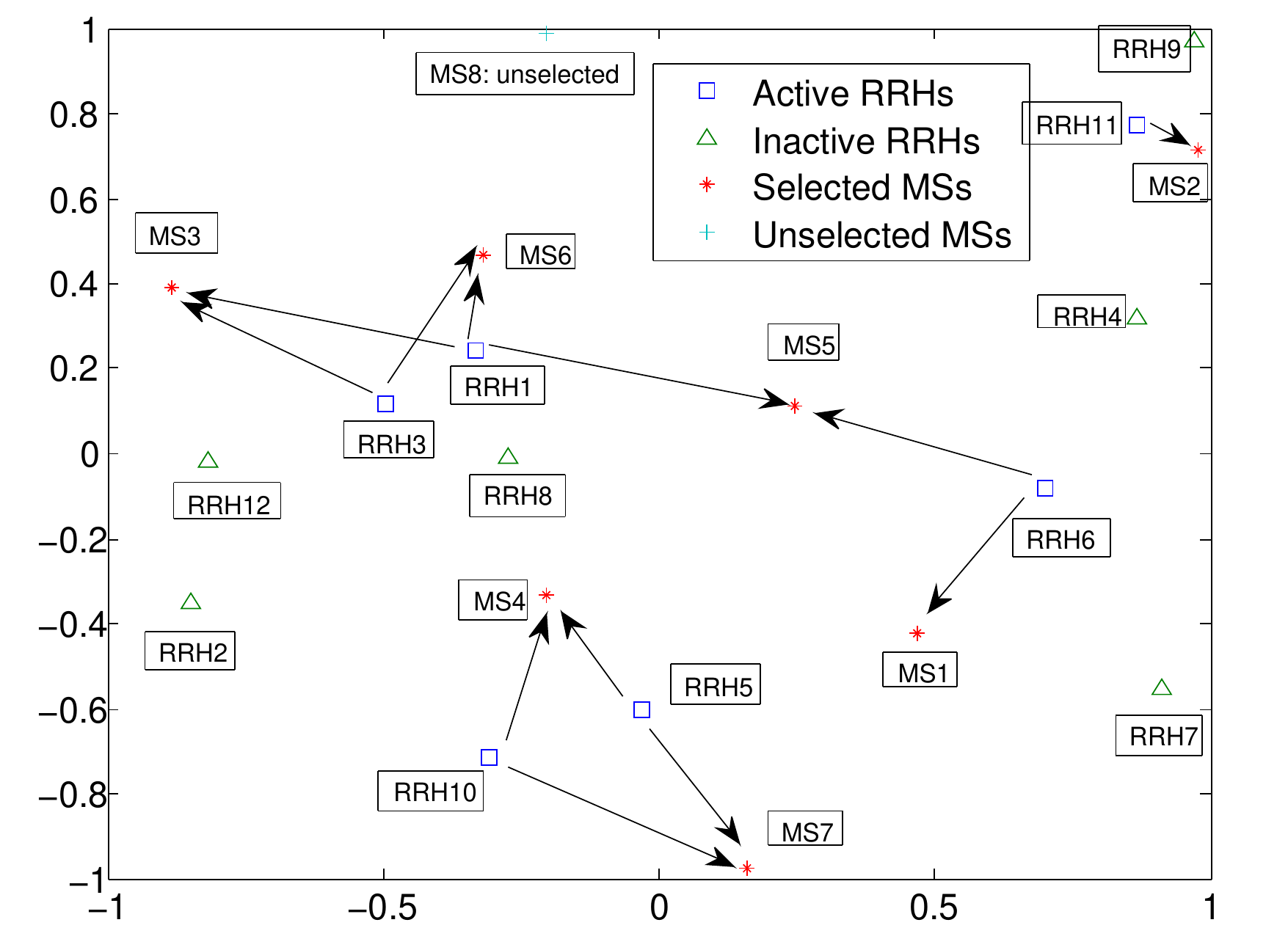}
\caption{The converged state of one randomly generated system configuration. The boundary user 8 is not selected as it is far from the RRHs.}
\label{convergedstate}
\end{minipage}%
\hfill
\begin{minipage}[t]{0.495\linewidth}
\centering
\includegraphics[width=2.7in]{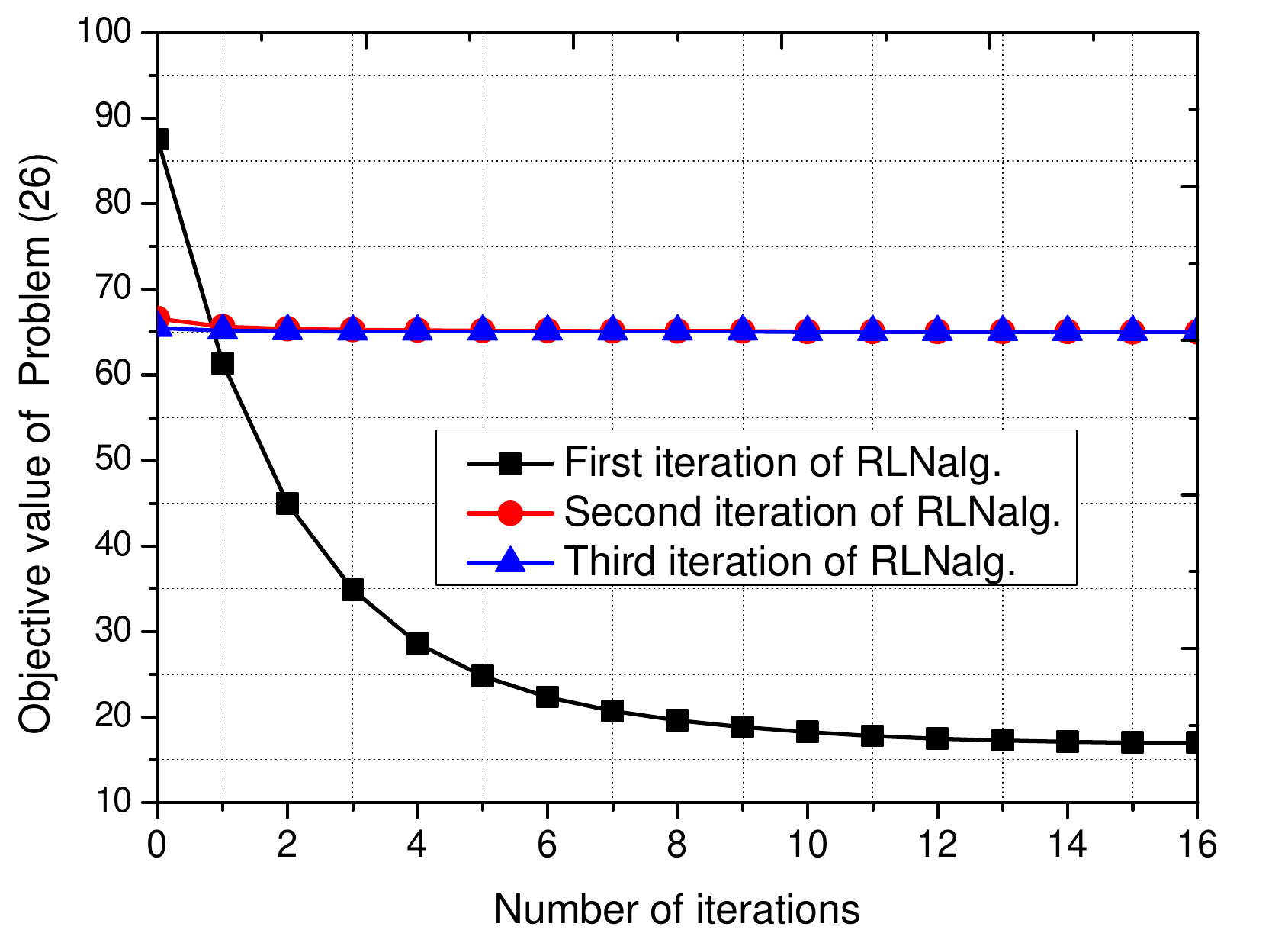}
\caption{The converged state of one randomly generated system configuration. The boundary user 8 is not selected as it is far from the RRHs.}
\label{convergeWMMSE}
\end{minipage}
\end{figure}

The convergence behaviours of the RLN algorithm are shown in  Figs.~\ref{RLNalgcon} (a) and (b) for the NPC and the number of the remaining RRHs in each iteration, respectively. Three different values of $\delta $ are tested, i.e., $\delta  = {10^{ - 4}},{10^{ - 5}}$ and ${10^{ - 6}}$.  One randomly generated channel is used to obtain the convergence behaviour, where the USC algorithm is first executed to find the largest feasible set of users. In this example, User 8 is removed to guarantee the feasibility of the other users as seen in Fig.~\ref{convergedstate}. It can be seen from the figures  that for all values of $\delta$, both the number of active RRHs and the NPC decrease rapidly and there is no additional decrease after the fifth iteration. At the converged state, only six RRHs are active. Compared to the full cooperation strategy where all RRHs are active, we can save large amount of power as seen from Fig.~\ref{RLNalgcon} (a). Fig.~\ref{convergedstate} illustrates the converged state of the system. It can be seen that RRH 2 is switched off since it is far from the users and User 8 is not selected as it is far from the RRHs. We also study the impact of initialization schemes on the performance of the RLN algorithm. The initial precoders for the RLN algorithm are the outputs of the USC algorithm which is initialized with the SVD-initial and rand-initial schemes. The simulation results show they achieve almost the same performance, which is not shown here for clarity.

\subsubsection{Convergence behaviour of the WMMSE algorithm}
\begin{figure}
\begin{minipage}[t]{0.495\linewidth}
\centering
\includegraphics[width=2.6in]{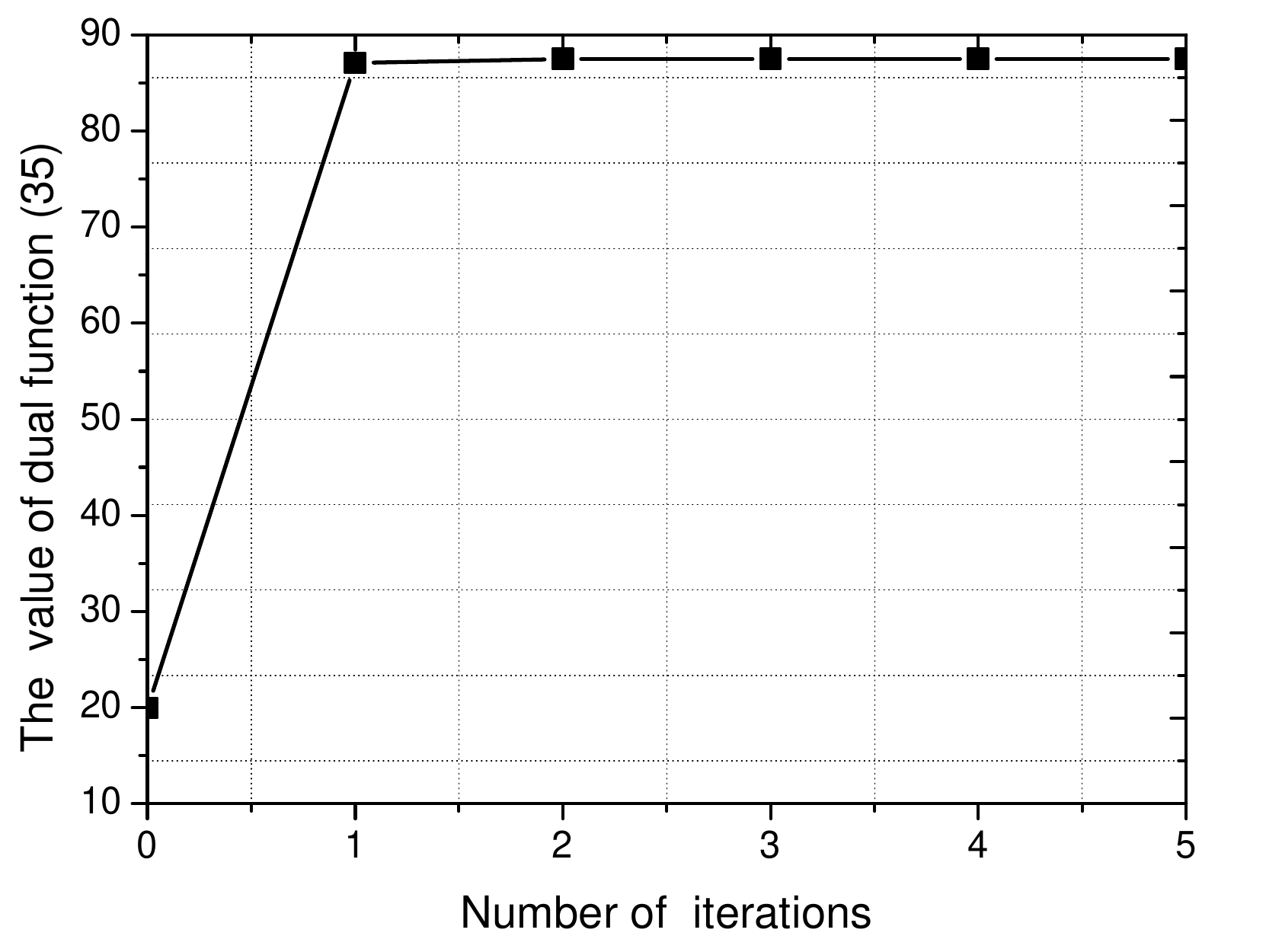}
\caption{Convergence behaviour of the BCD algorithm for the first iteration of the WMMSE algorithm.}
\label{BCDconver}
\end{minipage}%
\hfill
\begin{minipage}[t]{0.495\linewidth}
\centering
\includegraphics[width=2.8in]{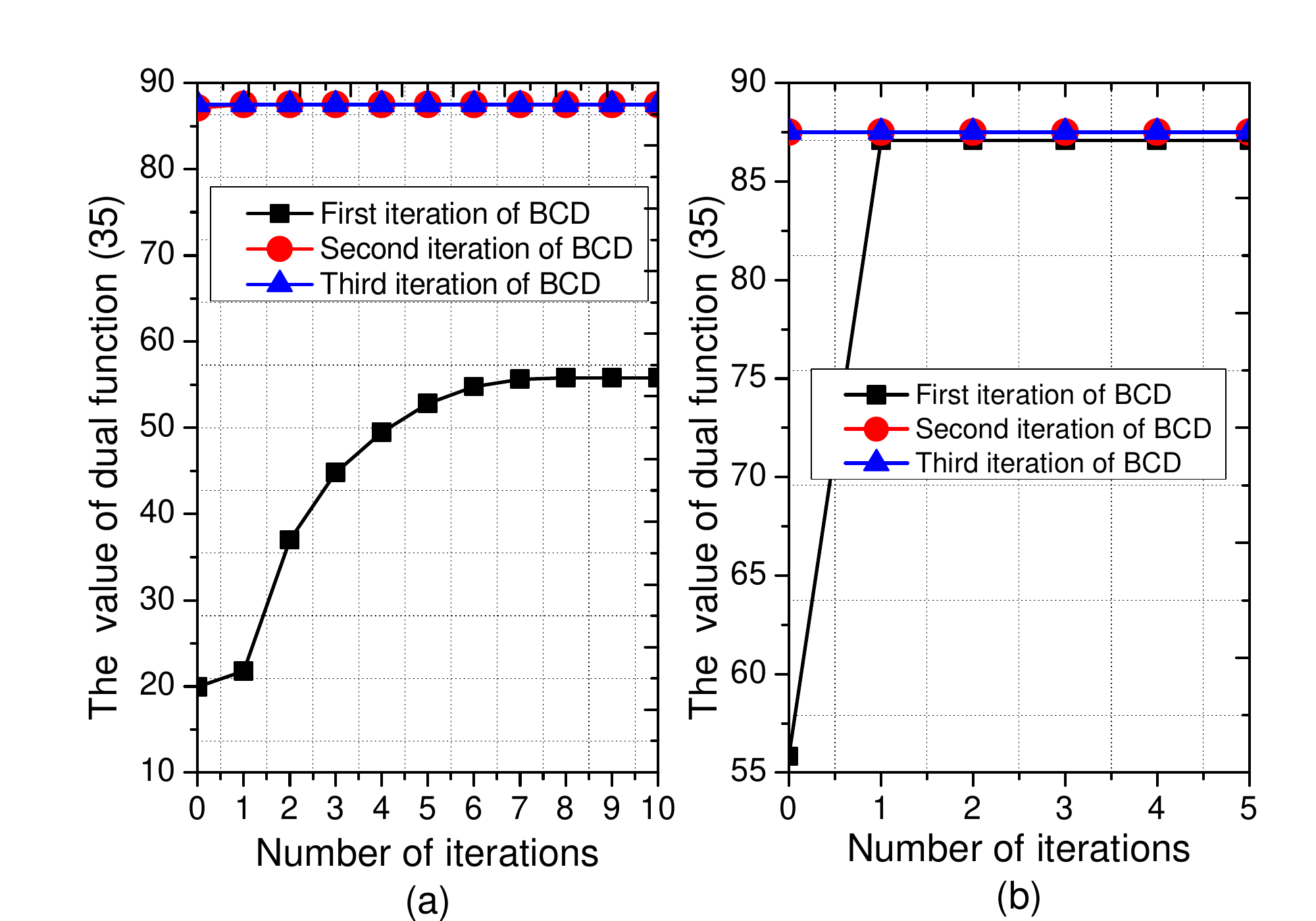}
\caption{(a)  Convergence behaviour of  Newton's method; (b) Convergence behaviour of  gradient descent method.}
\label{convernewton}
\end{minipage}
\end{figure}

In step 2 of each iteration of the RLN algorithm, we need to solve Problem (\ref{subproblem}) by using the WMMSE algorithm. Fig.~\ref{convergeWMMSE} shows  the convergence performance of the WMMSE algorithm for the first three iterations of the RLN algorithm. It is observed that the WMMSE algorithm converges within ten iterations for the first iteration of RLN algorithm. However, the objective values stay almost fixed  for the second and third iterations of the RLN algorithm.  This means that only in the first iteration of RLN algorithm, some iterations are required for the WMMSE algorithm.

\subsubsection{Convergence behaviour of the BCD algorithm}

In step 2 of each iteration of the WMMSE algorithm, Problem (\ref{equalprosub}) should be solved to update the precoding matrices by using the BCD algorithm. Fig.~\ref{BCDconver} shows the convergence behaviour of the BCD algorithm for the first iteration of the WMMSE algorithm. It is seen that the algorithm converges very fast and one iteration is sufficient to achieve a large portion of the converged value (99.2\% in this example).
\subsubsection{Convergence behaviour of  Newton's method and the gradient descent method}

In each iteration of the BCD algorithm, Newton's method is required to update $\left\{ {\lambda _k},\forall k \right\}$ and the gradient descent method is applied to update $\left\{ {\mu _i},\forall i \right\}$. The convergence behaviours of these two algorithms for the first three iterations of the BCD algorithm are shown in Figs.~\ref{convernewton} (a) and (b), respectively. Newton's method requires several iterations to converge only in the first iteration of the BCD algorithm, while stays almost constant for the second and third iterations of the BCD algorithm. Interestingly, the gradient descent method only requires one iteration to converge in the first iteration of the BCD algorithm and keeps fixed during the rest of the iterations of the BCD algorithm.  By combining the complexity analysis in (\ref{newtoncom}), (\ref{complexity}) and the above convergence behaviours, we can conclude that the BCD algorithm  has a much lower computational complexity than directly  solving the SOCP problem.

\subsubsection{Impacts of the number of data streams}
\begin{figure}
\begin{minipage}[t]{0.495\linewidth}
\centering
\includegraphics[width=2.4in]{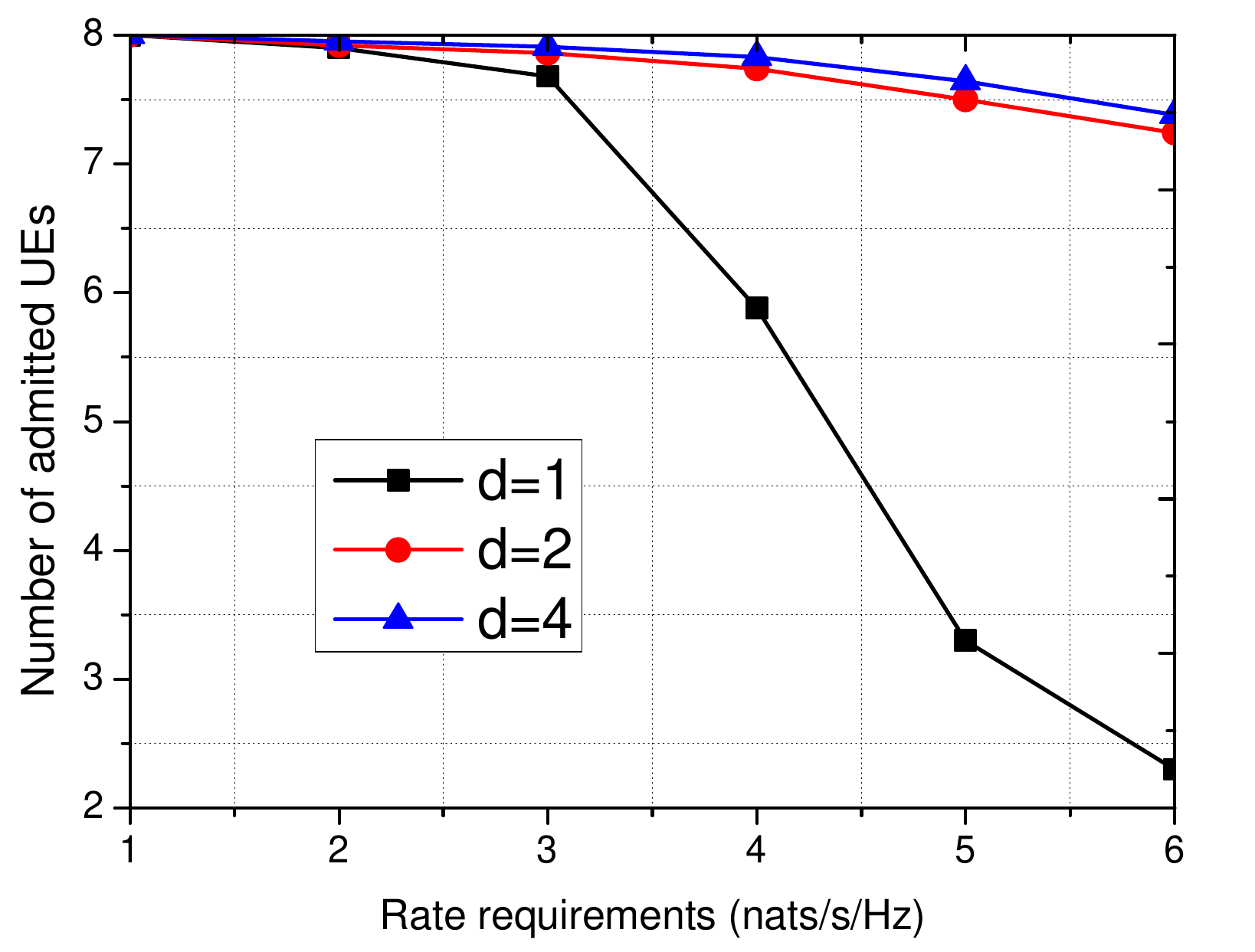}
\caption{The number of admitted users versus rate requirements for different numbers of data streams with $M=N=4$.}
\label{UEsnumberdata}
\end{minipage}%
\hfill
\begin{minipage}[t]{0.495\linewidth}
\centering
\includegraphics[width=2.8in]{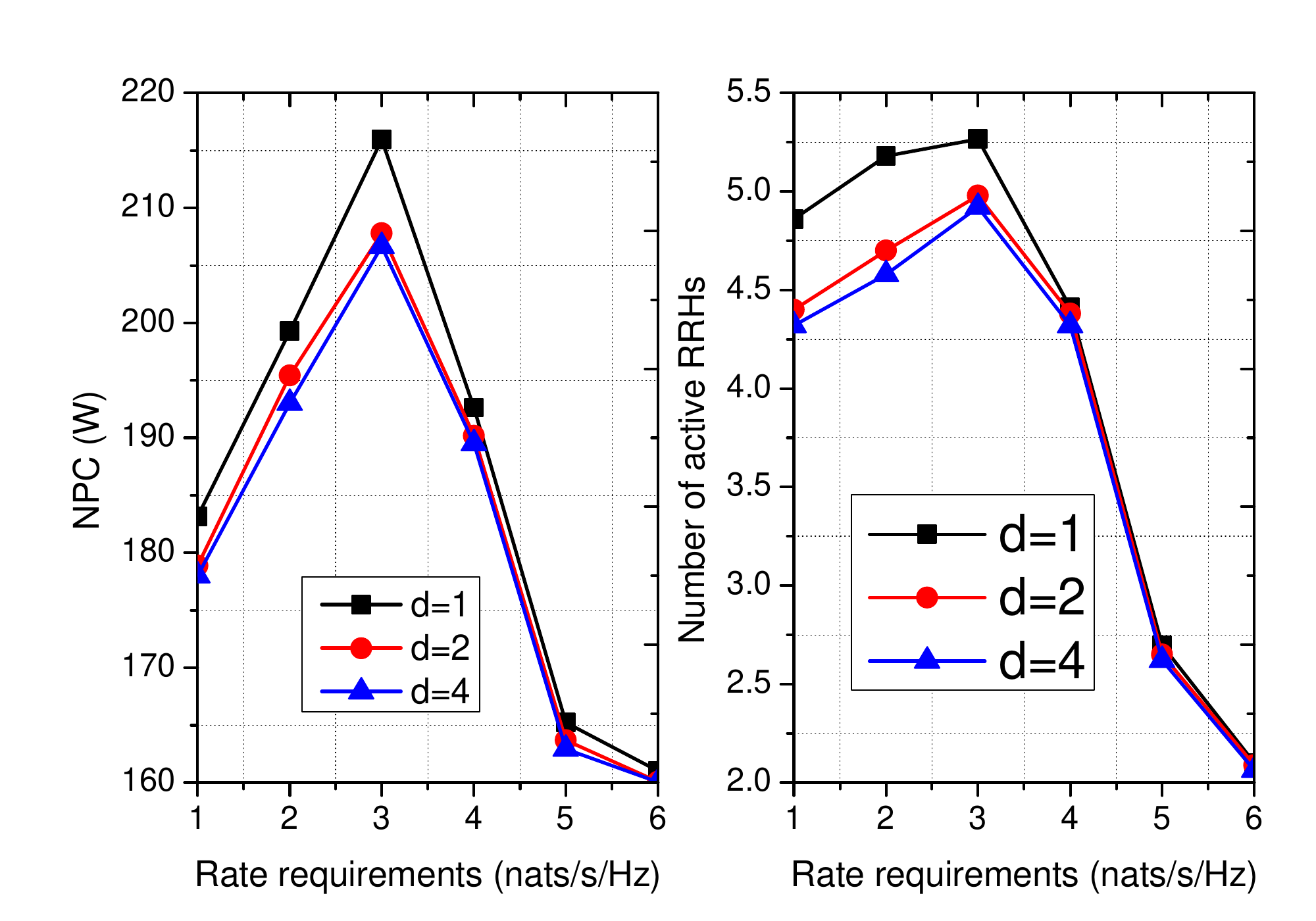}
\caption{(a) NPC versus the rate requirements; (b) The corresponding number of active RRHs versus the rate requirements.}
\label{Totalpownumberdata}
\end{minipage}
\end{figure}
In Fig.~\ref{UEsnumberdata}, the impact of the number of data streams on the number of admitted users is studied. As expected, the number of admitted users decreases with the rate requirements and larger number of data streams can support more users. We find significant performance gains can be achieved when the number of data streams increases from 1 to 2, especially for the high data rate requirements. However, only marginal performance gains are achieved by the case of $d=4$ over the case of $d=2$, which comes at the higher cost of computational complexity. This reveals that the performance saturates with the increase of data streams $d$. In Fig.~\ref{Totalpownumberdata}, the impacts of data streams on the NPC and on the number of active RRHs are studied with the same setup in Fig.~\ref{UEsnumberdata}. For fair comparison, we only consider the set of users that can be supported under the case of $d=1$, so that all cases can support the selected users. Fig.~\ref{Totalpownumberdata} (a) shows that the NPC first increases with the rate requirements when $R_{\rm{min}}\le 3 \ {\rm{nats/s/Hz}}$ and then decreases significantly when $R_{\rm{min}}>3 \ {\rm{nats/s/Hz}}$. The reason can be explained as follows. When $R_{\rm{min}}$ increases from $1 $ to $3 \ {\rm{nats/s/Hz}}$, the number of admitted users almost keeps stable as shown in Fig.~\ref{UEsnumberdata}, while the fronthaul power increases when the rate requirement increases and the number of active RRHs increases to support the higher rate requirements as seen in  Fig.~\ref{Totalpownumberdata} (b), which in turn consumes more power consumption. On the other hand, when $R_{\rm{min}} $ increase from $3$ to $6\ {\rm{nats/s/Hz}}$, the number of admitted users decreases dramatically as shown in Fig.~\ref{UEsnumberdata}, which leads to  reduced transmit power and a reduced number of active RRHs as shown in Fig.~\ref{Totalpownumberdata} (b). Again, it is observed from  Fig.~\ref{Totalpownumberdata} (a) that a greater number of data streams requires lower NPC, but the performance gain shrinks with the number of data streams.

\subsubsection{Impacts of the number of transmit antennas}
\begin{figure}
\begin{minipage}[t]{0.495\linewidth}
\centering
\includegraphics[width=2.4in]{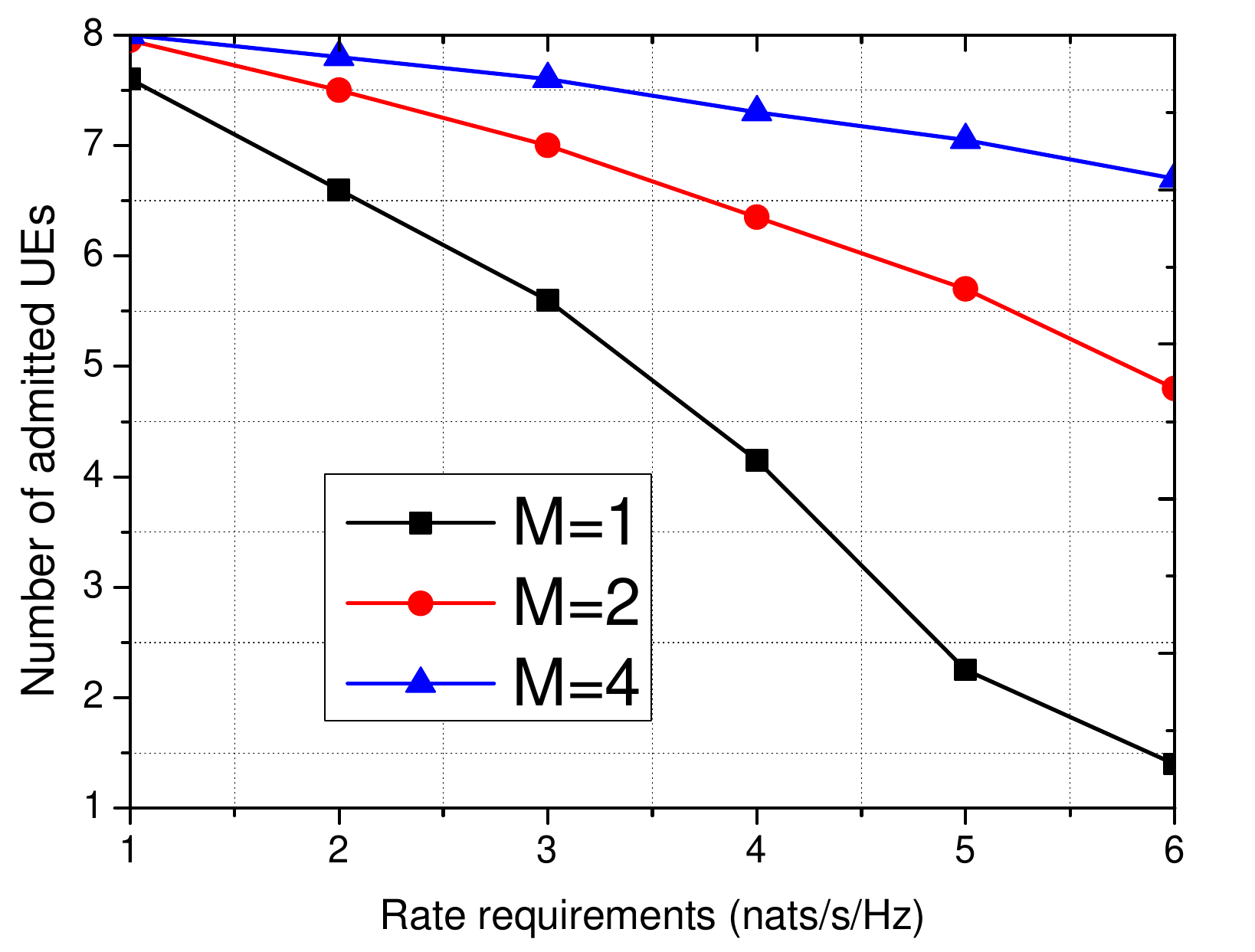}
\caption{The number of admitted users versus rate requirements for different numbers of transmit antennas with $N=2$.}
\label{UEsnumberantenna}
\end{minipage}%
\hfill
\begin{minipage}[t]{0.495\linewidth}
\centering
\includegraphics[width=2.8in]{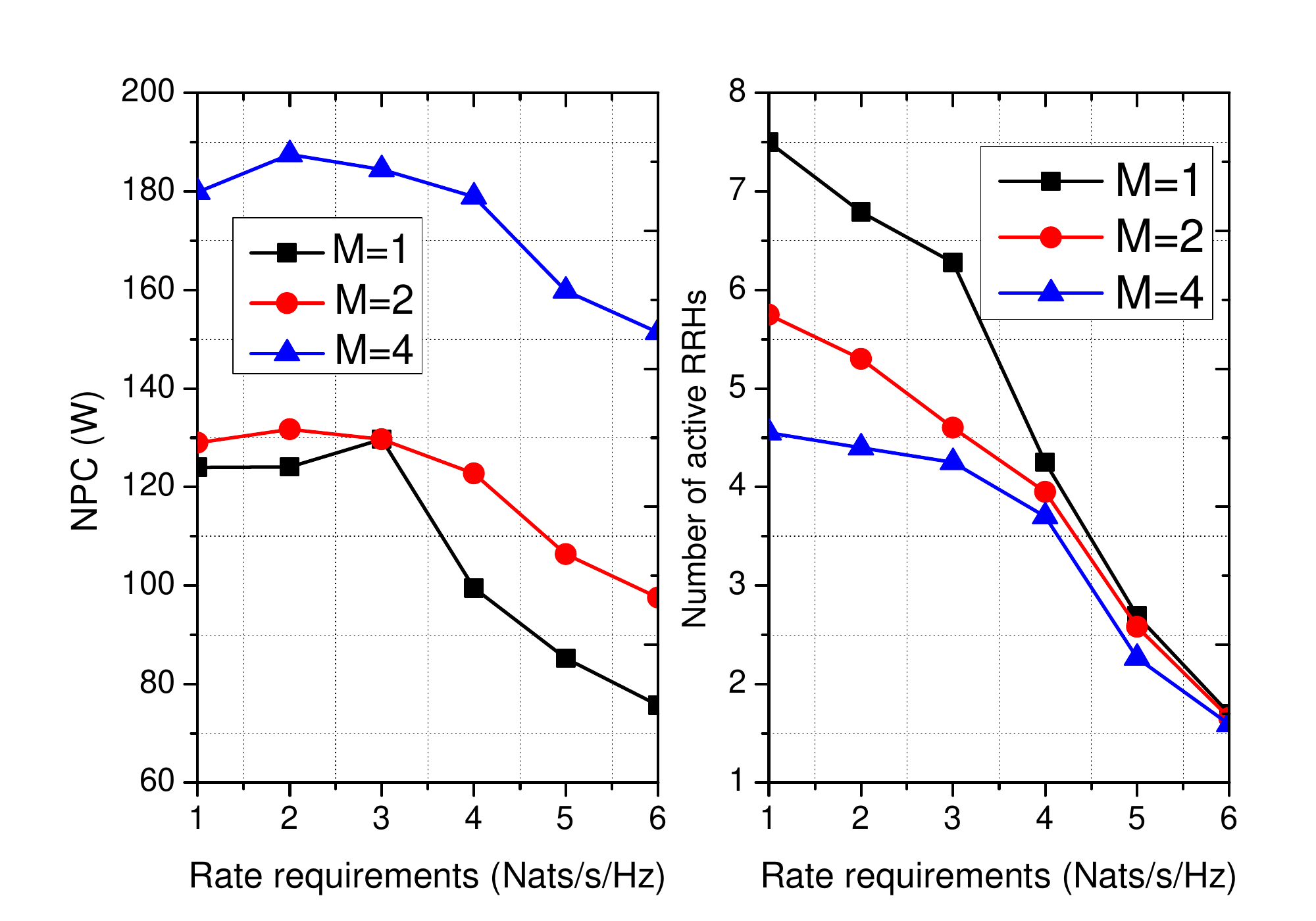}
\caption{(a) NPC versus the rate requirements; (b) The corresponding number of active RRHs versus the rate requirements.}
\label{Totalpownumberantenna}
\end{minipage}
\end{figure}
In Fig.~\ref{UEsnumberantenna}, the impact of the number of transmit antennas on the number of admitted users is studied. As expected, the number of admitted users increases with the number of transmit antennas due to more degrees of freedom. Significant performance gains can be achieved by the case of $M=2$ over the case of $M=1$, especially in the high rate regime. However, the performance gain shrinks for the case of $M=4$ over the case of $M=2$. In Fig.~\ref{Totalpownumberantenna}, the impacts of the number of antennas on the NPC and the number of active RRHs are investigated. For fair comparison, it is also assumed that the set of users selected from the case of $M=1$ are the input of Stage II for all cases of different values of $M$ so that  the selected users are the same and feasible for all cases. It is interesting to find that when $M$ increases, the NPC increases  while the number of active RRHs decreases. This is mainly due to the fact that the RRH power consumption model in (\ref{rrhpower}) increases linearly with $M$, and this increased power consumption dominates the reduced power consumption resulting from the reduced number of active RRHs. It should be emphasized that in some other cases with different values of system parameters, the NPC may not increase with $M$ and the counter part happens, such as the case of the low circuit power consumption for each antenna and high power consumption associated with the fronthaul power consumption.

\subsubsection{Impacts of the candidate size}

\begin{figure}
\begin{minipage}[t]{0.495\linewidth}
\centering
\includegraphics[width=2.4in]{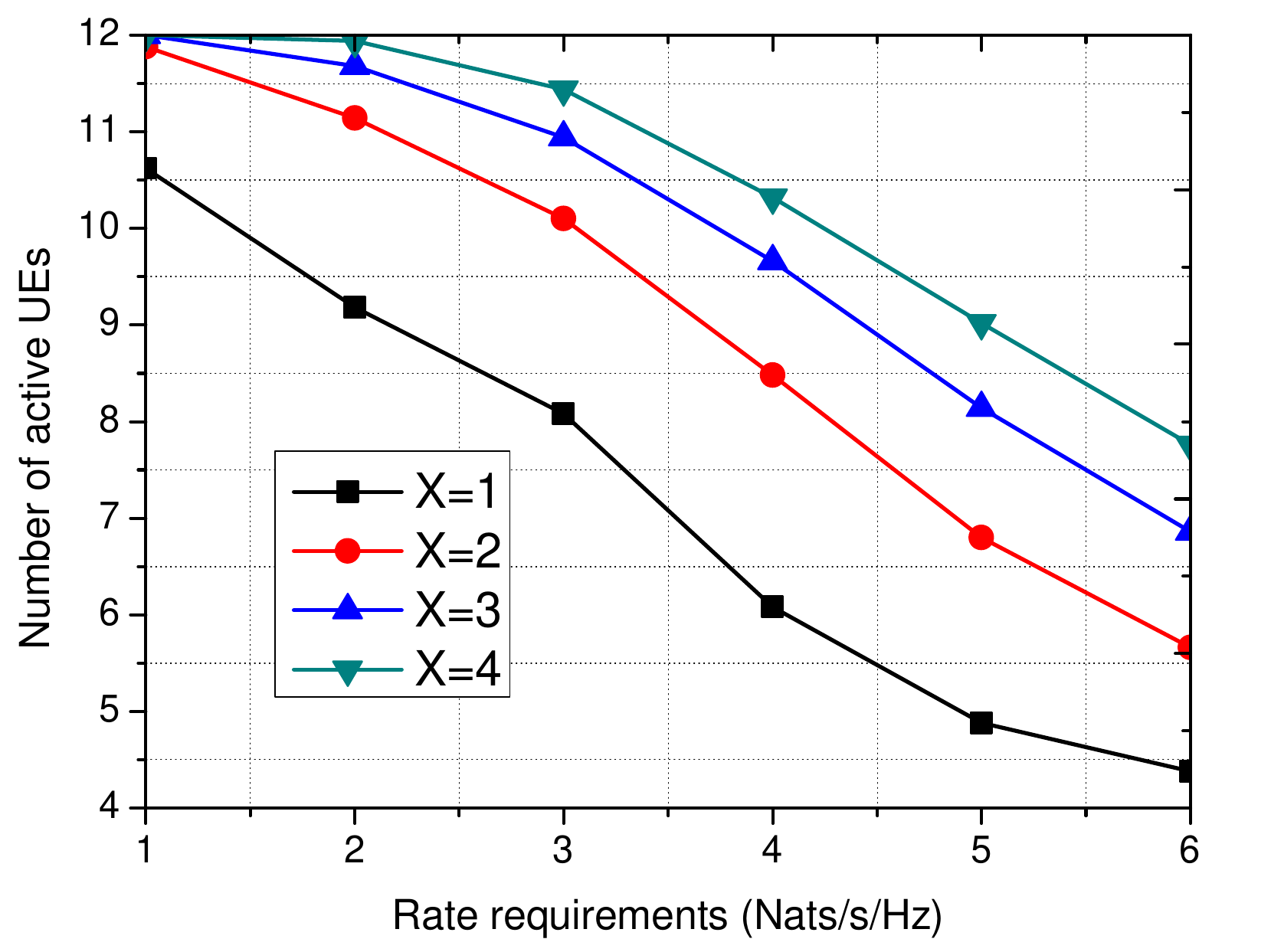}
\caption{The number of admitted users versus rate requirements for different candidate sizes.}
\label{UEsnumbercandi}
\end{minipage}%
\hfill
\begin{minipage}[t]{0.495\linewidth}
\centering
\includegraphics[width=2.8in]{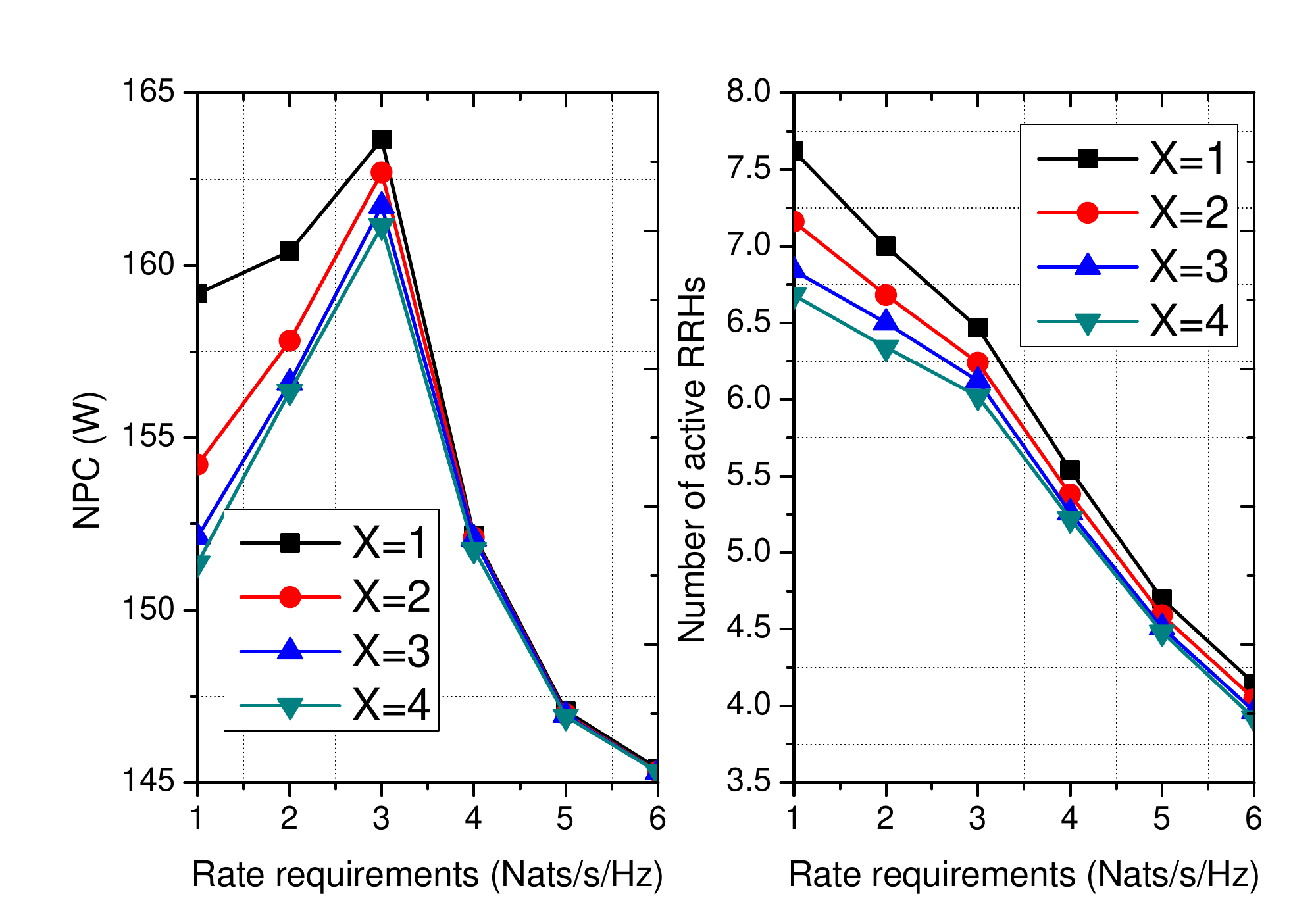}
\caption{(a) NPC versus the rate requirements; (b) The corresponding number of active RRHs versus the rate requirements.}
\label{Totalpowcandis}
\end{minipage}
\end{figure}
The impact of candidate size on the number of admitted users is illustrated in Fig.~\ref{UEsnumbercandi} for a dense network with 20 RRHs and 12 users. As expected, larger candidate sizes can support more users due to the increased degrees of freedom. However, the performance gains decreases with the candidate sizes, which implies that there is no need to consider distant RRHs for each user since they contribute less to their signal strength. In general, the candidate size should be no larger than 4 to achieve a good tradeoff between performance and complexity. Similarly to the trend observed in Fig.~\ref{UEsnumberdata} (a), it is seen from Fig.~\ref{Totalpowcandis} (a) that the NPC increases in the low rate regime, while decreasing significantly in the high rate regime. For the former part, the reason is that the increased fronthaul power dominates the reduced circuit power for the reduced active RRHs. While for the latter part, the reason is the opposite. Also, it is observed that the NPC performance gain for larger candidate size is more obvious in the low rate regime, while the performance is almost the same in the high rate regime. This is mainly due to the fact that in the high rate regime, only a small number of users can be admitted, and these users are separated far away. As a result, the multiuser interference is not so significant and each user's nearest RRH is able to serve it with the rate requirement.

\subsection{Performance comparison}

We compare the performance of the RLN algorithm with the following RRH selection methods:
 \begin{itemize}
   \item Exhaustive search (Exhau-search)  method: For each given active RRH set ${\cal A}$, this method first checks its feasibility. If feasible, the method will use the WMMSE algorithm to solve the corresponding transmit power minimization problem. The complexity of this method increase exponentially with $I$, which is served as the performance benchmark for our proposed algorithm.
   \item Successive RRH selection (Succesive-sel) method: This method first lets all the RRHs be active and check its feasibility. If  feasible,  the method applies the WMMSE algorithm to solve the transmit power minimization problem. Then, the method gradually removes the RRHs according to their transmit power from the lowest to the highest until the problem becomes infeasible. The complexity of this scheme increases linearly with $I$.
   \item Greedy search method: In each step, we exclude each RRH and calculate the NPC when the remaining RRHs are active. Then, we remove the RRH  so that the remaining RRHs yield the least NPC. This procedure terminates until the problem becomes infeasible. The complexity of this scheme increases quadratically with $I$.
   \item Full cooperative (Full-coop) method: In this method, all the selected RRHs in cluster-formation stage are active and the WMMSE algorithm is used to solve the transmit power minimization problem.
 \end{itemize}
 For fair comparison, we assume in the following simulation results, only the channel realizations that are feasible for all users are considered.
\subsubsection{Impact of the rate requirements}

\begin{figure}
\begin{minipage}[t]{0.495\linewidth}
\centering
\includegraphics[width=2.6in]{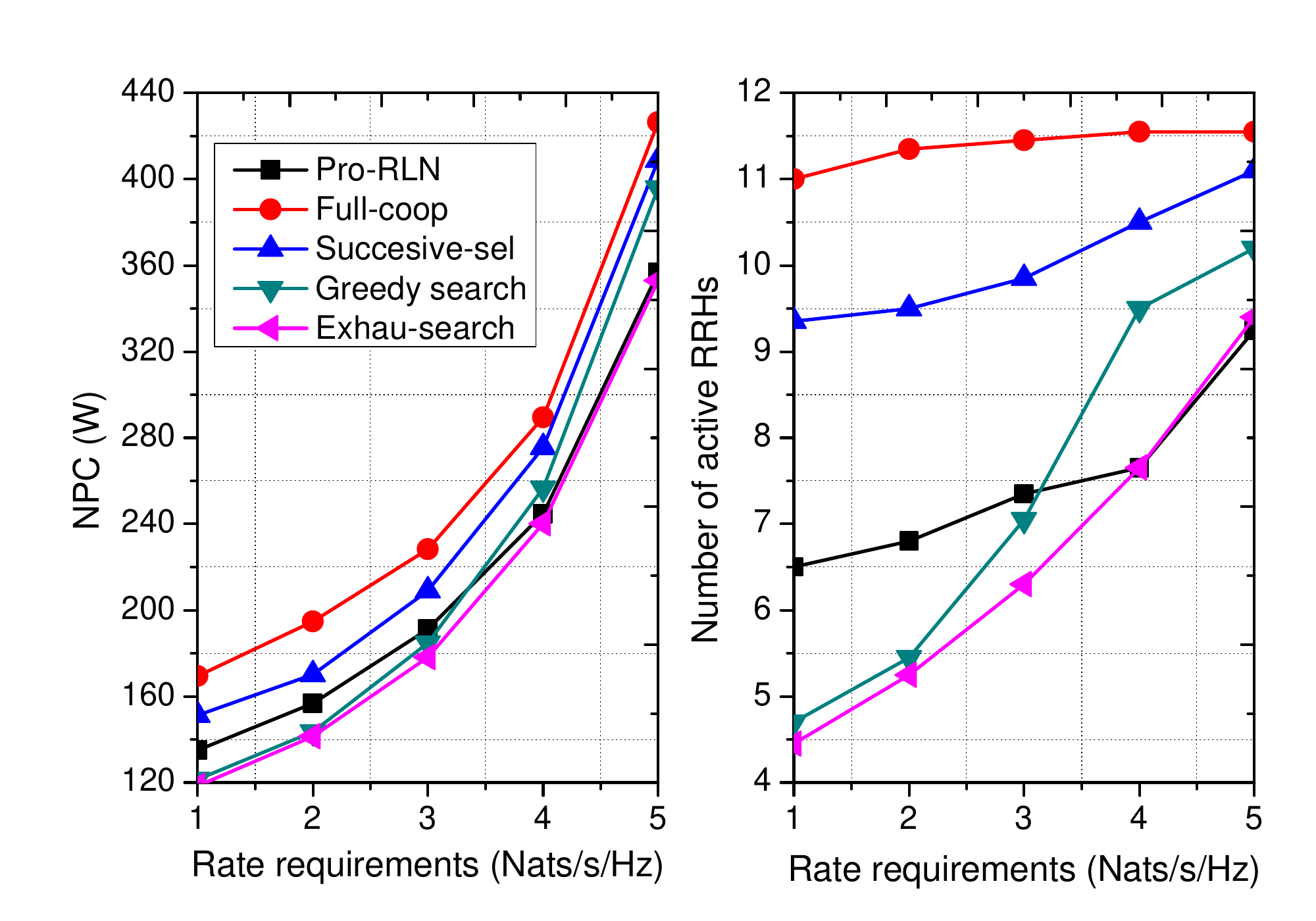}
\caption{(a) NPC versus the rate requirements; (b) The corresponding average number of active RRHs versus the rate requirements. The candidate size is $X=4$. }
\label{fig10}
\end{minipage}%
\hfill
\begin{minipage}[t]{0.495\linewidth}
\centering
\includegraphics[width=2.6in]{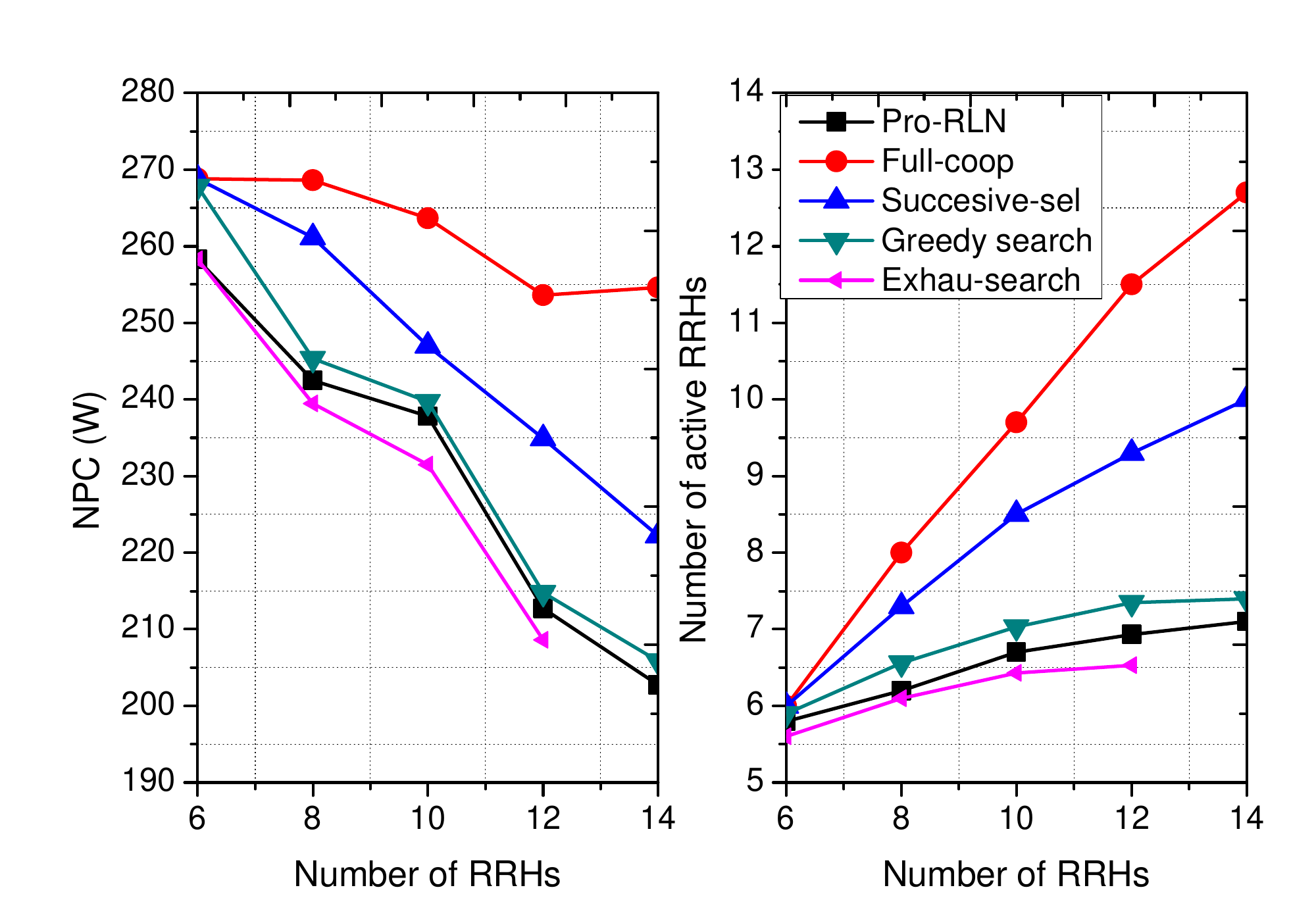}
\caption{(a) NPC versus the number of RRHs; (b) The corresponding number of active RRHs versus the number of RRHs with $R_{\rm{min}}=3\  {\rm{nats/s/Hz}}$ and $X=4$.}
\label{fig11}
\end{minipage}
\end{figure}

Figs.~\ref{fig10} (a) and (b) illustrate the average NPC and the corresponding  number of active RRHs versus the rate requirements, respectively. Fig.~\ref{fig10} (a) shows that the RLN algorithm outperforms the `Succesive-sel' method and `Full-coop' method for all rate regimes. However, the performance of the `Greedy search' method is slightly better than the RLN algorithm when $R_{\rm{min}}\le 3 {\rm{nats/s/Hz}}$, while the RLN algorithm outperforms the `Greedy search' method in the high rate regime and the performance gain increases with the rate requirements. Fig.~\ref{fig10} (b) shows a similar trend in terms of the number of active RRHs.  Compared with the optimal `Exhau-search' method, the performance loss in power consumption is at most $8\%$ when $R_{\rm{min}}=1$ nats/s/Hz, and this gap gradually diminishes with the increase of rate requirements. In particular, the performance gain provided by the `Exhau-search' method over the RLN algorithm is negligible when $R_{\rm{min}}=5$ nats/s/Hz. As expected, the  `Full-coop' method consumes the highest power since all selected RRHs are active.

\subsubsection{Impact of the number of RRHs}

Figs.~\ref{fig11} (a) and (b) illustrate the average NPC and the corresponding  number of active RRHs versus the total number of RRHs, respectively.  It is seen that the NPC achieved by all schemes decreases with $I$ due to the fact that when there are more RRHs, the average access distance between users and RRHs decreases significantly and thus leads to more reduced transmit power. It is again observed that the performance of the RLN algorithm is superior to that of the `Succesive-sel' method. This implies that selecting the RRHs only based on the transmit power is not enough, and may incur significant performance loss. Note that the `Greedy search' method requires higher power consumption than the RLN algorithm for all numbers of RRHs, especially when $I=6$. Also, the performance of `Exhau-search' method is slightly better than the RLN algorithm. Note that although the number of active RRHs increases slightly with the total number of  RRHs as seen in Figs.~\ref{fig11} (b), the NPC decreases. This may due to the fact that the overall transmit power reduction overwhelms the increase of circuit power.

\section{Conclusion}\label{conclusion}

In this paper, a joint selection of active RRHs and optimization of the precoding matrices which minimizes the NPC for the MIMO C-RAN, while guaranteeing users' rate requirements and per-RRH power constraints, has been studied. A low-complexity user selection was proposed to guarantee the feasibility of the other users. Then a low-complexity iterative algorithm, based on the reweighted $l_1$-norm minimization method, WMMSE algorithm, Newton's method, and gradient descent method, was proposed to solve the  network power minimization problem for the selected users. Simulation results show that the proposed algorithms converge fast, which is attractive for practical implementation. Also, more antennas at the user side can admit more users. The proposed user selection algorithm was shown to achieve the similar performance as the optimal exhaustive search method. Moreover, our proposed algorithm was shown to achieve much greater power savings than the full cooperation method, and the performance loss compared with the optimal approach is insignificant.

\numberwithin{equation}{section}
\begin{appendices}
\section{Proof of Theorem 1}\label{prooftheorem1}

In step 2 of the $n$th iteration, we solve Problem (\ref{socpfea}) to obtain the optimal $\{\alpha_k^{(n)}\}_{k\in \cal U}$ and ${\bf{V}}^{(n)}$ with given ${\bf{U}}^{(n-1)}$ and ${\bf{W}}^{(n-1)}$. Hence, we have ${h_k}\left( {{\bf{V}}^{(n)},{{\bf{U}}_k^{(n-1)}},{{\bf{W}}_k^{(n-1)}}} \right) \ge \left({\alpha_k^{(n)}}\right)^2 {R_{k,{\rm{min}}}},\forall k$. In step 3 of the $n$th iteration, we update ${\bf{U}}^{(n)}$ and ${\bf{W}}^{(n)}$ as in (\ref{receanu})  with ${\bf{V}}^{(n)}$. According to Lemma 1, we have $R_k\left( {{{\bf{V}}^{(n)}}} \right) = {h_k}\left( {{{\bf{V}}^{(n)}},{\bf{U}}_k^{(n)},{\bf{W}}_k^{(n)}} \right) \ge {h_k}\left( {{{\bf{V}}^{(n)}},{\bf{U}}_k^{(n - 1)},{\bf{W}}_k^{(n - 1)}} \right)$. Hence, we have
\begin{equation}\label{fesi}
  {h_k}\left( {{\bf{V}}^{(n)},{{\bf{U}}_k^{(n)}},{{\bf{W}}_k^{(n)}}} \right) \ge {\left(\alpha_k^{(n)}\right)}^2 {R_{k,{\rm{min}}}}.
\end{equation}
In step 2 of the $(n+1)$th iteration, we obtain $\{\alpha_k^{(n+1)}\}_{k\in \cal U}$ and ${\bf{V}}^{(n+1)}$ with given ${\bf{U}}^{(n)}$ and ${\bf{W}}^{(n)}$ by solving Problem (\ref{socpfea}). Then we have $\sum\nolimits_{k \in {\cal U}} {{{\left( {\alpha _k^{(n + 1)} - 1} \right)}^2}}  \le \sum\nolimits_{k \in {\cal U}} {{{\left( {\alpha _k^{(n)} - 1} \right)}^2}} $. The reason is that from (\ref{fesi}), $\{\alpha_k^{(n)}\}_{k\in \cal U}$ and ${\bf{V}}^{(n)}$ is just a feasible solution for Problem (\ref{socpfea}) with given ${\bf{U}}^{(n)}$ and ${\bf{W}}^{(n)}$. Hence, the objective value of Problem (\ref{alternativepro}) is monotonically decreasing. Obviously, the objective value is lower bounded by zero. Hence, Algorithm \ref{iterdada} will converge.

\section{Proof of Theorem 2}\label{prooftheorem2}

We first prove that the sequence of ${\bf{V}}$ generated by the WMMSE algorithm (i.e. Algorithm \ref{wmmse}) always satisfies the rate requirements of Problem (\ref{subproblem}). In step 2, we obtain ${\bf{V}}^{(l)}$ with ${\bf{U}}^{(l-1)}$ and ${\bf{W}}^{(l-1)}$. Hence, ${h_k}\left( {{\bf{V}}^{(l)},{\bf{U}}_k^{(l-1)},{\bf{W}}_k^{(l-1)}} \right) \ge {R_{k,{\text{min}}}},\forall k$
 hold since ${\bf{V}}^{(l)}$ is feasible for Problem (\ref{Equivalentpro}). According to Lemma 1, ${h_k}\left( {{\bf{V}}^{(l)},{\bf{U}}_k^{(l-1)},{\bf{W}}_k^{(l-1)}} \right)$ is a lower-bound of ${R_k}({\bf{V}}^{(l)})$, i.e., ${R_k}({\bf{V}}^{(l)})\geq {h_k}\left( {{\bf{V}}^{(l)},{\bf{U}}_k^{(l-1)},{\bf{W}}_k^{(l-1)}} \right)$. Hence, ${R_k}({\bf{V}}^{(l)})\geq R_{k,{\text{min}}}$ holds. Thus, the sequence of ${\bf{V}}$ generated by the WMMSE algorithm satisfies the rate requirements of Problem (\ref{subproblem}).

Next, we show that the value of the objective function of Problem (\ref{subproblem}) monotonically decreases during the iterative process of the WMMSE algorithm. Denote ${\rm{Obj}}({\bf{V}}^{(l)})$ as the objective value of Problem (\ref{subproblem}) when ${\bf{V}}={\bf{V}}^{(l)}$.  Step 2 of the WMMSE algorithm updates ${\bf{V}}^{(l)}$ by solving Problem (\ref{equalprosub}) with ${\bf{U}}^{(l-1)}$ and ${\bf{W}}^{(l-1)}$. The objective value of this step, ${\rm{Obj}}({\bf{V}}^{(l)})$,  will be no larger than ${\rm{Obj}}({\bf{V}}^{(l-1)})$, i.e., ${\rm{Obj}}({\bf{V}}^{(l)})\leq {\rm{Obj}}({\bf{V}}^{(l-1)})$. The reason is that  ${\bf{V}}^{(l-1)}$ is a feasible solution for Problem (\ref{equalprosub}) with ${\bf{U}}_k^{(l-1)}$ and ${\bf{W}}_k^{(l-1)}$ since ${h_k}\left( {{\bf{V}}^{(l-1)},{\bf{U}}_k^{(l-1)},{\bf{W}}_k^{(l-1)}} \right)={R_k}({\bf{V}}^{(l-1)}) \ge {R_{k,{\text{min}}}}$ holds as proved above. In step 3 of the WMMSE algorithm, we update ${\bf{U}}^{(l)}$ and ${\bf{W}}^{(l)}$ by using (\ref{receanu})  with ${\bf{V}}^{(l)}$. This step increases the value of ${h_k}\left( {{\bf{V}},{\bf{U}}_k,{\bf{W}}_k} \right)$ while maintaining the same objective value of Problem (\ref{subproblem}). Therefore, this step provides ``room'' for the next iteration to decrease the objective value. In addition, the objective value is lower bounded by zero. Hence, the WMMSE algorithm converges.

Then, we prove that given the initial set of precoders, the WMMSE algorithm converges to a unique solution. Obviously, when ${\bf{V}}$ is given, ${\bf{U}}$ and ${\bf{W}}$ can be uniquely determined by (\ref{receanu}). The remaining task is to prove that given ${\bf{U}}$ and ${\bf{W}}$, the BCD algorithm can obtain the unique globally optimal solution ${\bf{V}}$. Since $\{{\bf{G}}_k,\forall k\}$ are  positive definite matrices, the objective function in Problem (\ref{equalprosub}) is a strictly convex function with respect to (w.r.t.) $\bf{V}$. Obviously, the constraints in Problem (\ref{equalprosub})  are convex w.r.t. $\bf{V}$ \cite{boyd2004convex}. Hence, Problem (\ref{equalprosub}) is a strictly convex problem \cite{boyd2004convex}. According to [Page 137 in \cite{boyd2004convex}], the globally optimal solution of Problem (\ref{equalprosub}) is unique. On the other hand, Theorem 3 proves that the BCD algorithm can obtain the globally optimal solution to the dual problem (\ref{dualproblem}). As  Problem (\ref{equalprosub}) is a convex problem and it satisfies the Slater's condition \cite{boyd2004convex}, the duality gap between Problem (\ref{equalprosub}) and its dual problem (\ref{dualproblem}) is zero \cite{boyd2004convex}. As a result, the BCD algorithm can obtain the unique globally optimal solution ${\bf{V}}$. Finally, by alternatively updating step 2 and step 3, the WMMSE algorithm will converge to a unique solution. It should be emphasized that as Problem (\ref{Equivalentpro}) is non-convex, it may have many locally optimal solutions, and  the unique solution of the WMMSE algorithm depends on the initial point. However, given the initial points of precoders, the WMMSE algorithm will converge to a unique solution.

Finally, we prove that the unique solution satisfies the KKT conditions of Problem (\ref{subproblem}). Denote the converged solution of the WMMSE algorithm as ${\bf{V}}^{\star}$, ${\bf{U}}^{\star}$ and ${\bf{W}}^{\star}$. With given ${\bf{U}}^{\star}$ and ${\bf{W}}^{\star}$, the Lagrange function of Problem  (\ref{Equivalentpro})  can be written as
\begin{eqnarray}
{\cal L}\left( {{\bf{V}},{\bm{\lambda}} ,{\bm{\mu}} } \right) &=& \sum\nolimits_{k \in {\cal U}} {{\bf{\bar V}}_k^H} {{\bf{G}}_k}{{{\bf{\bar V}}}_k} + \sum\nolimits_{k\in \cal U} {{\lambda _k}\left( {R_{k,{\rm{min}}}}- {h_k}\left( {{\bf{V}},{\bf{U}}_k^{\star},{\bf{W}}_k^{\star}} \right)  \right)}\nonumber\\
&+& \sum\nolimits_{i \in \cal I} {{\mu _i}\left( \sum\nolimits_{k \in {{\cal U}_i}} {\left\| {{{\bf{B}}_{i,k}}{{{\bf{\bar V}}}_k}} \right\|} _F^2 - {P_{i,\max }}   \right)},\label{laglanri}
\end{eqnarray}
where ${{\bm{\lambda}} } = \left\{ {\lambda _k, \forall k\in \cal U} \right\}$ and ${\bm{\mu} } = \left\{ {\mu _i,\forall i\in \cal I } \right\}$ are the corresponding  Lagrange multipliers.

According to Theorem 3,  the BCD algorithm can obtain the globally optimal solution of Problem (\ref{equalprosub}) (also Problem (\ref{Equivalentpro})) with given ${\bf{U}}^{\star}$ and ${\bf{W}}^{\star}$, there must exist ${{\bm{\lambda}} ^{\star}}$ and ${\bm{\mu} ^{\star}}$ such that $\{  {\bf{V}}^{\star},{\bm{\lambda}}^{\star},\bm{\mu}^{\star}\}$ satisfy the following KKT conditions
 \begin{eqnarray}
{\nabla _{{{\bf{\bar V}}_k}}}{\cal L} &=& \nabla _{{{\bf{\bar V}}_k}} \sum\nolimits_{k \in {\cal U}} {{\bf{\bar V}}_k^{\star,H}} {{\bf{G}}_k}{{{\bf{\bar V}}}_{k}^{\star}}- \sum\nolimits_{k \in \cal U} {\lambda _k^{\star}{\nabla _{{{\bf{\bar V}}_k}}}{h_k}\left( {{\bf{V}}^{\star},{\bf{U}}_k^{\star},{\bf{W}}_k^{\star}} \right))}\nonumber \\
&&+ \sum\nolimits_{i\in \cal I} {\mu _i^{\star}} {\nabla _{{{\bf{\bar V}}_k}}}\left( \sum\nolimits_{k \in {{\cal U}_i}} {\left\| {{{\bf{B}}_{i,k}}{{{\bf{\bar V}}}_k^{\star}}} \right\|} _F^2  \right)= {\bf{0}},\forall k\in \cal U,\label{lav} \\
&&\lambda _k^{\star}\left( {{h_k}\left( {{\bf{V}}^{\star},{\bf{U}}_k^{\star},{\bf{W}}_k^{\star}} \right) - {R_{k,{\rm{min}}}}} \right) = 0,\forall k\in \cal U,\label{raterecp}\\
&&\mu _i^{\star}\left( P_{i,\max } - \sum\nolimits_{k \in {{\cal U}_i}} {\left\| {{{\bf{B}}_{i,k}}{{{\bf{\bar V}}}_k^{\star}}} \right\|} _F^2\right) = 0,\forall i\in \cal I,\label{powercp}\\
&&{h_k}\left( {{\bf{V}}^{\star},{\bf{U}}_k^{\star},{\bf{W}}_k^{\star}} \right) \ge {R_{k,{\rm{min}}}},\forall k\in \cal U,\label{ratere}\\
&& \sum\nolimits_{k \in {{\cal U}_i}} {\left\| {{{\bf{B}}_{i,k}}{{{\bf{\bar V}}}_k^{\star}}} \right\|} _F^2 \le {P_{i,\max }},\forall i\in \cal I.\label{powerre}
\end{eqnarray}

Since ${\bf{U}}^{\star}$ and ${\bf{W}}^{\star}$ are updated by using (\ref{receanu}), we have ${h_k}\left( {{\bf{V}}^{\star},{\bf{U}}_k^{\star},{\bf{W}}_k^{\star}} \right)={R_k}({\bf{V}}^{\star})$ according to Lemma 1. By substituting it into the equations (\ref{lav}), (\ref{raterecp}) and (\ref{ratere}), we find that the set of equations (\ref{lav})-(\ref{powerre}) are just the KKT conditions of Problem (\ref{subproblem}).

\section{Proof of Theorem 3}\label{prooftheorem3}

According to \cite{boyd2004convex}, the dual problem of any optimization problem is a convex problem. Thus, the dual problem (\ref{dualproblem}) is jointly convex with respect to ${\bm{\lambda}}$ and $\bm{\mu}$. Assuming that the constraint of this problem satisfies the Slater's condition, the KKT condition of this problem is sufficient and necessary for optimality. For given $\bm{\mu}$, the dual problem (\ref{dualproblem}) is a convex problem w.r.t. ${\bm{\lambda}}$. According to \cite{boyd2004convex}, Newton's method can obtain the globally optimal solution of dual problem (\ref{dualproblem}) for given $\bm{\mu}$. In addition, for given ${\bm{\lambda}}$, the dual problem (\ref{dualproblem}) is convex w.r.t. $\bm{\mu}$, and the gradient descent method can be applied to obtain the globally optimal solution. Then by adopting the same idea as in the proof of Theorem 1 in \cite{weiyu}, we can prove that the converged solution also satisfies the KKT condition of Problem (\ref{dualproblem}). Since Problem (\ref{dualproblem}) is a convex optimization problem, Algorithm 6 can attain the globally optimal solution of Problem (\ref{dualproblem}).

\end{appendices}



\
\






\vspace{-0.9cm}
\bibliographystyle{IEEEtran}
\bibliography{myre}


\end{document}